\documentclass[12pt]{article}
\usepackage{graphicx}
\usepackage{amsmath}
\usepackage{amssymb}
\usepackage{bbm}

\begin{document}

\begin{titlepage}

\begin{center}

\begin{flushright}
CERN-TH/2003-271
 \\[12ex]
\end{flushright}

\textbf{\large Electromagnetic fluxes, monopoles, \\
and the order of the $4d$ compact U(1) phase transition.}
\\[6ex]

{Michele Vettorazzo$^{a,}$\footnote{vettoraz@phys.ethz.ch} and
Philippe de Forcrand$^{a,b,}$\footnote{forcrand@phys.ethz.ch}}
\\[6ex]
{${}^a${\it Institute for Theoretical Physics, ETH Z\"urich,
CH-8093 Z\"urich, Switzerland}\\[1ex]
${}^b${\it CERN, Theory Division, CH-1211 Gen\`{e}ve 23, Switzerland}}
\\[10ex]
{\small \bf Abstract}\\[2ex]
\begin{minipage}{14cm}{\small
We consider the $4d$ compact $U(1)$ gauge theory with extended action

\begin{equation}
S=-\beta \sum_P \cos \theta_{P} - \gamma \sum_P \cos 2\theta_{P}
\nonumber
\end{equation}

We give a full characterization of the phase diagram of this model using the notion of
\emph{flux}. The relation with the usual monopole picture is discussed. In analogy with the
XY model we consider the \emph{helicity modulus} \cite{Jose:1977gm} for this theory, and
show that it is an order parameter. Analyzing the finite-size effects of the helicity
modulus we conclude that the transition is first-order. The value of this order parameter
is related to the renormalized coupling $\beta_{R}$. We
measure $\beta^c_{R}$ at the transition point and give a counterexample to its conjectured universal value
\cite{Cardy:jg}.}
\end{minipage}
\end{center}
\vspace{1cm}

\end{titlepage}

\section{Introduction}
\label{sec:introduction}

A phase transition in an infinite system is rigorously defined by
a non-analyticity of the free energy as a function of the driving
parameter (like the temperature, an external field or some
coupling of the theory). In practice, a useful description can be
provided by identifying an observable (the so-called \emph{order
parameter}) whose vacuum expectation value is zero in one phase
and non-zero in the other. This behavior is manifestly
non-analytic and can sometimes be used to relate the phase
transition with the spontaneous breaking of some symmetry
(implicitly or explicitly defined) of the model, thus providing an
appealing physical picture of the transition itself.

The formulation of an order parameter is a highly non-trivial problem, and closely depends
on the specific features of the theory considered. A broad class of theories (for a review
see \cite{Froehlich}), which includes the $4d$ compact Abelian theory we are going to
consider, admits a characterization of the transition in terms  of \emph{topological
defects}, which is a general name used to indicate some collective degrees of freedom with
non-trivial homotopy group in terms of which one tries to reformulate the partition
function of the system: the picture which emerges is a theory of interacting defects
(sometimes one says that there is a \emph{gas of defects}), which are very dilute in one
phase and are condensed in the other. The nature of such a description is truly
non-perturbative.

An open question is the following: is the choice of the excitation
responsible for the phase transition unique? or is there some
freedom in the identification of the topological structure
involved? Here, we consider this question for the compact $U(1)$
system on a hypertorus.

In $3d$, a convenient choice of defect is the pointlike
\emph{magnetic monopole} \cite{DeGrand:eq}; in $4d$ we naturally
have, instead of pointlike objects, \emph{monopole currents}
representing monopoles evolving in space-time. The partition
function of the theory can be expressed in terms of these currents
\cite{Banks:1977cc} and the following picture emerges: in addition
to a weak-coupling Coulomb phase in which only small and dilute
loops are present, one finds a strong-coupling confined phase in
which there are many monopole loops deeply entangled.

Suppose now that one monitors the value of the electromagnetic
(e.m.) \emph{flux} through some plane; we expect to observe two
completely different situations in the two regimes: the flux
should be constant if no monopoles are present, while it should
dramatically vary in the other phase. This is the physical picture
we are going to consider, namely a picture of the phase transition
in terms of fluxes and not in terms of monopoles: they are
equivalent from the logical point of view; there can then be
arguments in favor of one or the other picture, but none
`precedes' the other in a strict logical sense.

As we will extensively see, this alternative point of view
provides a way to define a natural order parameter for the system,
the so-called \emph{helicity modulus}, defined by the response of
the system to a variation of the flux. An accurate determination
of this observable will be used as the basis for a finite-size
effect analysis, which will provide good evidence about the
first-order nature of the transition.\\

The paper is organized in the following way. In sections \ref{sec:the_model} and
\ref{sec:flux_definition} we review the model under study and the notion  of flux on a
lattice. In section \ref{sec:character_phase_trans} we show how the behavior of fluxes
qualitatively changes across the phase boundaries of the system, and thus provide a first
characterization of the model in terms of fluxes. We also interpret this behavior in terms
of the monopole picture, showing the equivalence of the two approaches (we will not review
the monopole picture \cite{DeGrand:eq} \cite{Banks:1977cc}, but only use some results to
interpret our data). In section \ref{sec:Charact_of_phases} we develop an analogy between
our Abelian context and the non-Abelian one, proposing the analog of 't Hooft's twisted
boundary conditions \cite{'tHooft:1979uj}. In section \ref{sec:The helicity modulus} we
introduce an order parameter for our theory, the helicity modulus \cite{Nelson}. In section
\ref{sec:Helicity_modulus_numerical} we show our numerical results for the order parameter
in the whole phase diagram; in the following three sub-sections we discuss $(i)$ the issue
of the order of the phase transition, $(ii)$ the relation between the helicity modulus and
the renormalized coupling (in the Coulomb phase), and $(iii)$ the conjecture (first
proposed by Cardy \cite{Cardy:jg}) about the universality of its value at the transition.
In section \ref{sec:C-periodic_b_c} we show how the presence of a monopole in the lattice
can be enforced by adding a flux and using C-periodic boundary conditions; the free energy
of a monopole is studied and a comparison with other constructions is presented.
Conclusions follow.
Preliminary results about sections 2-5 have been presented in \cite{deForcrand:2002ym}. 
In Table \ref{tab:simulation_road_map} in the Appendix we present a summary of the simulations 
performed for  the paper.

\section{The model}
\label{sec:the_model}

Consider a $4d$ lattice with periodic boundary conditions
(p.b.c.). Call $a$ the lattice spacing. We associate to links a
$U(1)$ field $U_\mu(r)=\exp(i\theta_\mu(r))=\exp(iae A_\mu(r))$,
where $e$ is the gauge coupling, $r=a x_\mu \hat e_\mu$
($x_\mu=1,\ldots,L_\mu$) and $\hat e_\mu$ is the unit vector in
direction $\mu=1,\ldots,4$. We work with the following action:

\begin{equation}
\label{eq:the_model} S=-\beta \sum_P \cos \theta_{P} - \gamma
\sum_P \cos 2\theta_{P}
\end{equation}

\noindent
where $\theta_{P}\equiv \theta_{\mu\nu}(r)=\theta_\mu(r)+\theta_\nu(r+\hat
e_\mu)-\theta_\mu(r+\hat e_\nu)-\theta_\nu(r)$ is the plaquette angle, $\beta$ is the
Wilson coupling, and we refer to $\gamma$ as the \emph{extended} coupling. This model was
introduced by Bhanot \cite{Bhanot:1981zg} and has attracted interest also recently
\cite{Campos:1998jp}.

In order to point out the symmetries of the model consider the
following transformation of link variables

\begin{eqnarray}
U_1(x,y,z,t)  &\to& U_1(x,y,z,t)        \nonumber \\
U_2(x,y,z,t)  &\to& \left\{ \begin{array}{ll}
                            \tilde U\cdot U_2(x,y,z,t) & \mbox{if $x$
                            is odd}\\
                            U_2(x,y,z,t)          & \mbox{if $x$ is
                            even}
                            \end{array}
                            \right. \
                                        \nonumber  \\
U_3(x,y,z,t)  &\to& \left\{ \begin{array}{ll}
                            \tilde U\cdot U_3(x,y,z,t) & \mbox{if $x+y$
                            is odd}\\
                            U_3(x,y,z,t)          & \mbox{if $x+y$ is
                            even}
                            \end{array}
                            \right. \
                                        \nonumber  \\
U_4(x,y,z,t)  &\to& \left\{ \begin{array}{ll}
                            \tilde U\cdot U_4(x,y,z,t) & \mbox{if $x+y+z$
                            is odd}\\
                            U_4(x,y,z,t)          & \mbox{if $x+y+z$ is
                            even}
                            \end{array}
                            \right. \
\end{eqnarray}

\noindent
where $\tilde U$ is a $U(1)$ matrix at will. This
transformation is such that

\begin{itemize}
\item if $\tilde U=-1=e^{i\pi}$ every configuration with couplings
$(\beta,\gamma)$ is mapped into one with couplings
$(-\beta,\gamma)$. But since this is just a change of variables,
we conclude that the mapping $\beta \to -\beta$ is a symmetry of
the theory.

\item if $\tilde U=e^{i\frac{\pi}{2}}$ we map $(0,\gamma) \to
(0,-\gamma)$, which is therefore another symmetry.

\end{itemize}

\noindent
Furthermore, the change of variables $\theta_\mu(x) \to
2\theta_\mu(x)$ maps the couplings $(\beta,0)$ onto $(0,\beta)$,
showing the equivalence of the model on the two coordinate axes.

\noindent
Finally, the limit $\gamma \to +\infty$ corresponds to
recovering the $Z_2$ gauge theory ($\theta_P=\lbrace 0,\pi
\rbrace$; a
gauge-transformation then fixes the links to $Z_2$).\\

\begin{figure}[h]
\begin{center}
\includegraphics[height=6.0cm]{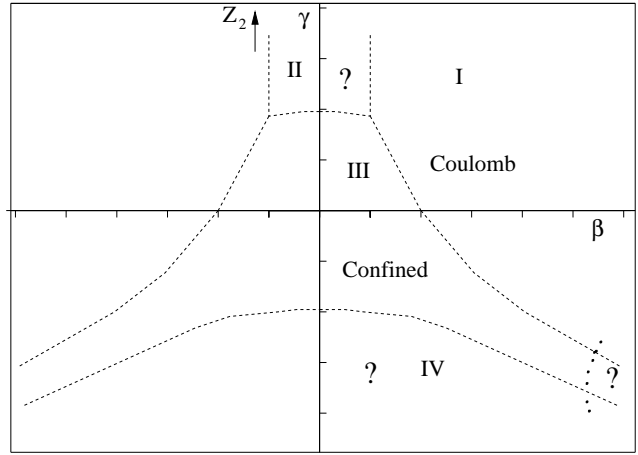}
\end{center}
\caption{\label{fig:phase_diagram}Phase diagram of the extended
$U(1)$ model.}
\end{figure}

In Fig.~\ref{fig:phase_diagram} we draw a sketch of the phase
diagram.

Three unknown aspects of the phase diagram are indicated with
question marks:

\begin{enumerate}
\item  the nature of the phase indicated with $II$; \item  the
nature of the phase $IV$; \item  the fate of the two phase
boundaries in the bottom right corner of the phase diagram (the
same holds for the bottom left corner).
\end{enumerate}

Another long standing question, still not completely resolved, is the order of the phase
transition, especially along the $I-III$ line. A recent paper \cite{Arnold:2002jk} shows,
through a high statistics analysis of the plaquette action distribution, that the
transition is first-order along the Wilson axis. We will confirm this result using a
totally different observable. It is also numerically known that the transition becomes
weaker and weaker as $\gamma$ becomes negative, which motivated the hypothesis that a
tricritical point could exist \cite{Evertz:1984pn} along the phase boundary between regions
$I$ and $III$ (see Fig.~\ref{fig:phase_diagram}), beyond which the transition could be
second-order. We will also address this issue with our new observable.\\

We now introduce the notion of flux, which will provide the tool
we need to explore the phase diagram.

\section{The flux in Abelian gauge theories}
\label{sec:flux_definition}

To introduce the notion of flux we need in $4d$, it is more
natural to start from the $2d$ case. Consider a compact $U(1)$
gauge theory defined on a $2d$ lattice, with sizes $L_\mu,L_\nu$,
and p.b.c.. Let us consider the definition

\begin{equation}
\label{eq:2d_flux} \Phi=\sum_P
\hspace{0.05cm}[\theta_{P}]_{-\pi,\pi}
\end{equation}

\noindent
where $[\theta_P]_{-\pi,\pi}$ is the plaquette angle
reduced to the interval $(-\pi,\pi)$ (its so-called $physical$
part), and the sum is over all plaquettes. This quantity, due to
p.b.c., is $2\pi k$ valued ($k\in Z$), because each link is summed
twice with opposite signs, giving, overall, $0$ modulo $2\pi$.

If we consider the limit $a \to 0$, we obtain the standard result

\begin{equation}
\frac{1}{ea^2}(\theta_P)_{\mu\nu} \to F_{\mu\nu}=\partial_\mu
A_\nu -\partial_\nu A_\mu
\end{equation}

\noindent
where $F_{\mu\nu}$ is the field strength; so the
definition in Eq.~(\ref{eq:2d_flux}) becomes

\begin{equation}
\frac{1}{e}\Phi_{\mu\nu}=\int F_{\mu\nu} d\sigma
\end{equation}

\noindent
which motivates the interpretation of $\Phi$ as an
\emph{e.m. flux}. Furthermore, in this limit we notice that fluxes
corresponding to different values of $k$ are \emph{topologically}
disconnected, in the sense that no tunneling is possible between
two flux values through a smooth variation of gauge field. In
fact, to change the value of the flux, one plaquette angle must
exceed the value of $\pi$; the statistical suppression of this
change is $e^{-\beta\Delta S} \sim e^{-2\beta}$, which goes to
zero in the continuum limit ($\beta \to \infty$). \footnote{In the
$2d$ case the lack of ergodicity through flux values of
simulations at large $\beta$ can be
cured by a cluster algorithm \cite{Sinclair:vm}.}\\

Consider now the same theory, but with $d=4$. Call $L_\mu,
L_\nu,L_\sigma$ and $L_\rho$ the four sizes of the lattice. Our
definition of flux through any $(\mu,\nu)$ orientation is:

\begin{equation}
\label{eq:lattice_flux} \Phi_{\mu\nu}=\frac{1}{L_\rho
L_\sigma}\sum_{\mu\nu \hspace{0.1cm} planes}
\sum_{P_{\mu\nu}}\hspace{0.05cm}[(\theta_P)_{\mu\nu}]_{-\pi,\pi}
\end{equation}

\noindent
A double sum is present:

\begin{itemize}
\item   the \emph{internal} $\sum_{P_{\mu\nu}}$ is the sum over the
plaquettes in a single plane. So plane by plane we use the same
definition as in the $2d$ case.

\item  the \emph{external} average $\frac{1}{L_\rho
L_\sigma}\sum_{\mu\nu \hspace{0.1cm} planes}$ over all parallel
planes of a given orientation is non-trivial because the flux
through different planes can change (in $3d$, for example, there
are monopoles between planes). The allowed values for
$\Phi_{\mu\nu}$ are thus multiples of $2\pi/L_\rho L_\sigma$.
\end{itemize}

Configurations whose flux is $2\pi k$ play a special role, for
reasons which will be clear in the next section, and we say that
they define \emph{flux sectors}.

Let us compute the action of a flux sector in the \emph{classical}
(minimal action) limit. In this case the flux through a single
plane is equally distributed over all plaquettes and it does not
change across parallel planes:

\begin{equation}
\label{eq:plaq_in_classical_limit}
(\theta_P)_{\mu\nu}=\frac{\Phi}{L_\mu L_\nu}=\frac{2\pi k}{L_\mu
L_\nu}, \quad k\in Z
\end{equation}

\noindent
We obtain (here for the Wilson action)

\begin{equation}
\label{eq:classical_flux_action} S_k=2\beta \pi^2 k^2\frac{L_\rho
L_\sigma}{L_\mu L_\nu}
\end{equation}

\noindent
showing that higher values of the flux are strongly
suppressed. We stress that in the classical limit, even if
$\Phi/2\pi$ is not integer valued because of the presence, e.g.,
of an external field, the action is anyway quadratic in the flux:

\begin{equation}
\label{eq:action_quadr_in_flux} S=\frac{1}{2}\beta\Phi^2
\frac{L_\rho L_\sigma}{L_\mu L_\nu}
\end{equation}

\noindent
In all actual simulations we will work with hypercubic
lattices, where $L_\mu = L \hspace{0.2cm}\forall \mu$, so that the
geometric factor $\frac{L_\rho L_\sigma}{L_\mu L_\nu}$ is $1$.


\begin{figure}[h]
\begin{center}
\includegraphics[height=10.0cm,angle=-90]{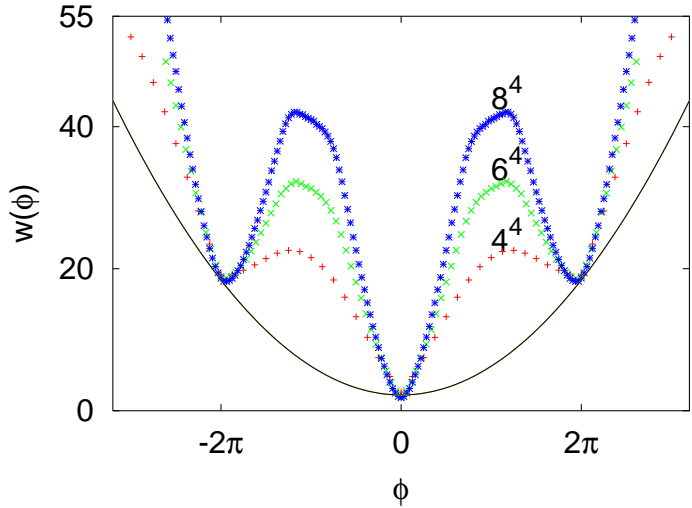}
\end{center}
\caption{\label{fig:figure4}Flux effective potential in the
Coulomb phase at $\beta=1.2$ and  $\gamma=0$ (for lattice size
$L=4,6,8$). The dashed parabolic line is
Eq.~(\ref{eq:action_quadr_in_flux}) with $\beta=0.85$; for a
discussion of this value see section \ref{sec:the
_rinormalized_coupling}.}
\end{figure}

\begin{figure}[h]
\begin{center}
\includegraphics[height=10.0cm,angle=-90]{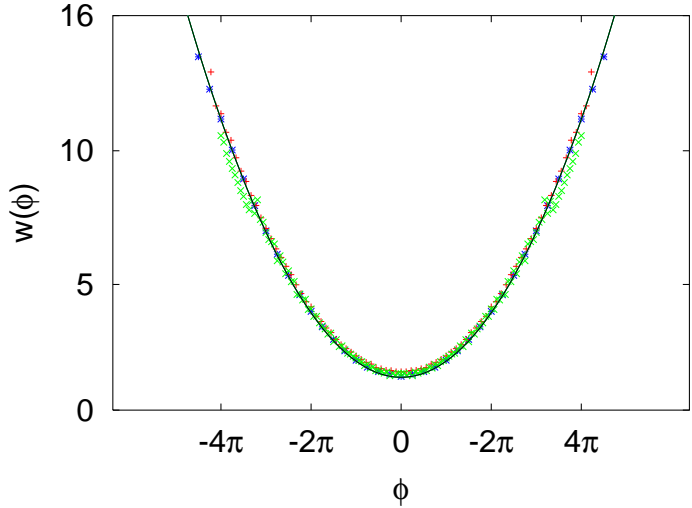}
\end{center}
\caption{\label{fig:figure5} Flux effective potential in the
confined phase at $\beta=0.8$ and  $\gamma=0$ ($L=4,6,8$). A
parabolic interpolating curve is superimposed
($\rho(\phi)=e^{-\phi^2/2\sigma^2}$, $\sigma=2.8$).}
\end{figure}

\section{Characterization of the phase transition}
\label{sec:character_phase_trans}

In this section we  consider the behavior of flux sectors across a
phase boundary. For this purpose we measure the \emph{flux
distribution} $\rho(\phi)$ through one $(\mu,\nu)$ orientation
during a Monte Carlo simulation. Since this quantity extends over
several orders of magnitude, we will show the \emph{flux effective
potential} $w(\phi)$, defined by

\begin{equation}
\label{eq:flux_effective_potential} \rho(\phi)=e^{-w(\phi)}
\end{equation}

\noindent
The results of our simulations are the following:

\begin{enumerate}

\item  In the Coulomb phase we observe minima of the flux
effective potential at flux values $2\pi k ~(k\in Z)$
(Fig.~\ref{fig:figure4}). As the thermodynamic limit is approached
we observe barriers of diverging height between those minima,
while the flux effective potential at $2\pi k$ does not change.
Extrapolating to the thermodynamic limit, we obtain that no
tunneling is allowed between sectors: this property finally
motivates the use of the word `sector' we introduced at the
beginning.

\item  In the confined phase (Fig.~\ref{fig:figure5}) sectors are
not defined, in the sense that no local minima of the flux
potential are observed. The distribution of the flux values is
continuous (in the limit $L \to \infty$), gaussian and independent
of the lattice size.
\end{enumerate}

\noindent
From the point of view of the flux picture of the model,
Figures \ref{fig:figure4} and \ref{fig:figure5} characterize
completely the phase transition.

\begin{figure}[h]
\begin{center}
\includegraphics[height=4.0cm,angle=0]{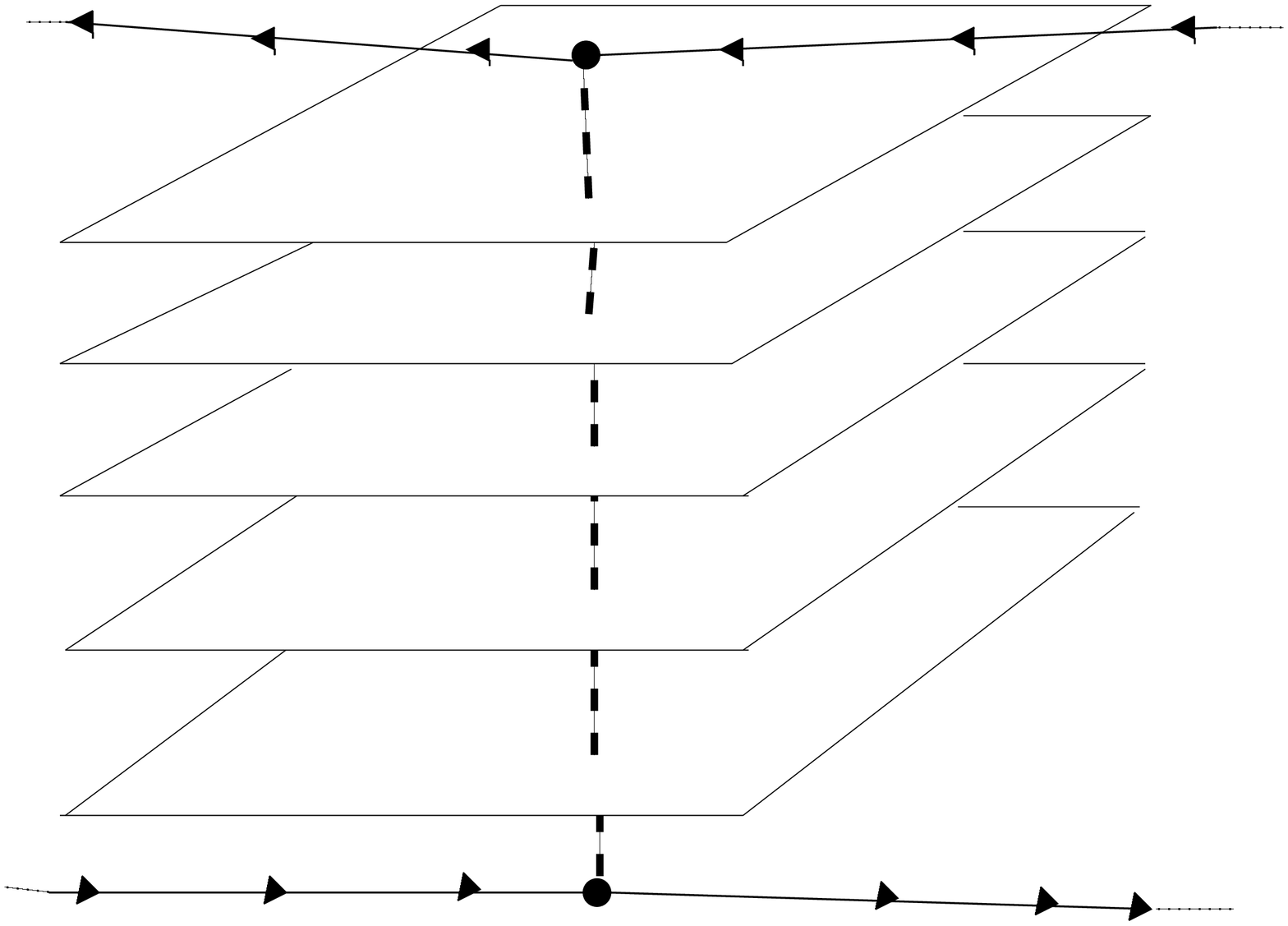}
\end{center}
\caption{\label{fig:dirac_sheet} Monopole configuration
corresponding to a flux tunneling event. The dashed line
represents the Dirac string, the solid lines the monopole
world-lines, wrapping around the boundaries.}
\end{figure}

We now want to interpret all these results in terms of the
monopole picture. We first discuss the interpretation of a
\emph{flux tunneling event} (in the Coulomb phase, when
$L<\infty$, we observed that tunnelings across flux sectors are
not forbidden, only suppressed). Let us start from the observation
that monopoles are created in pairs and that a plane pierced by a
Dirac string \cite{DeGrand:eq} gets an extra flux of $2\pi$.
Consider a monopole configuration of the kind shown in
Fig.~\ref{fig:dirac_sheet} where a monopole-antimonopole pair at a
certain distance $d$ wraps around the periodic boundaries; in
Fig.~\ref{fig:dirac_sheet} we show that a stack of $d$ planes is
pierced by a Dirac string. Let $d\to L_\mu$ (the lattice size):
due to p.b.c. the two particles annihilate, and the configuration
realized is the so-called \emph{Dirac sheet}. This is the
configuration responsible for a tunneling event; any other
configuration produces a variation of flux only in a fraction of
the planes, and thus is responsible for the non-integer ($mod
~2\pi$) values of the flux.

In the Coulomb phase only a few, small loops are created, so it is
very unlikely to have monopole lines wrapping around the
boundaries. If the volume is finite, there is a finite probability
$p$ for this event to happen, which vanishes like $e^{-L}$ as
$L\to \infty$, as one can observe in Fig.~\ref{fig:figure4}. The
reason for this exponential decay is that a tunneling event is
possible only in presence of two monopole loops of length $L$
(wrapping loops), and the free energy cost of a loop is
proportional to its length. \\

In the confined phase, instead, a large number of monopoles and
percolating loops of arbitrarily large size are present, so that
the flux values through different planes are totally
\emph{de-correlated}; the flux through one plane can be considered
as a random variable with a given distribution. The same
definition applied to another plane gives another
\emph{independent} random variable, with the same distribution.
Therefore the total flux Eq.~(\ref{eq:lattice_flux}) is just the
normalized sum of independent random variables, whose distribution
is gaussian as $L \to \infty$. It is remarkable that also the
smallest lattice we used ($L=4$, that is, with $16$ parallel
planes in the $\mu,\nu$ orientation) has the same distribution as
the largest one, indicating that the central limit `converges'
quickly to its asymptotic distribution.

Moreover, the variance $\sigma$ of the gaussian distribution is
related to the density $\delta$ of monopole currents. Call $f$ the
fraction of monopoles which contribute to flux disorder. Then,
between two successive planes, $L^2 f \delta $ monopoles
contribute to flux change. If the global flux (averaged over $L^2$
planes) has variance $\sigma^2$, the flux through one plane has
variance $L^2 \sigma^2$ and the difference of flux between two
planes has variance $2 L^2 \sigma^2$. In this way we estimate

\begin{equation}
f \delta = 2 (\frac{\sigma}{2\pi})^2
\end{equation}

\noindent
where we divide $\sigma$ by $2\pi$ to relate the units of flux ($\in 2\pi N$) to
the number of monopoles ($\in N$) through which we define $\delta$. Knowing $\sigma$ (see
caption of Fig.~\ref{fig:figure5}) and measuring $\delta=0.488$, we estimate $f \sim 0.81$:
the physical picture is thus that $80\%$ of the monopoles contribute to flux disorder at
$\beta=0.8$. We checked that this result does not
significantly change if we consider stronger couplings.\\


We now briefly describe the algorithm we used. The measurement of $\rho(\phi)$
(Eq.~(\ref{eq:flux_effective_potential})) is almost impossible using an ordinary local
updating algorithm, because of the strong suppression of higher flux values we mentioned at
the end of section \ref{sec:flux_definition}. To overcome this problem, we implemented a
\emph{multicanonical algorithm} \cite{Berg:cf}: iteratively, starting from an initial
distribution of flux values over a region $R_0=\lbrace \phi \mid
-\phi_0<\phi<\phi_0\rbrace$, we biased the sampling probability of the flux $\phi$ such
that we could extend the region over which $\rho(\phi)$ is known while maintaining a flat
flux distribution in the middle. This bias is removed during the statistical averaging.

\begin{figure}[h]
\begin{center}
\includegraphics[height=4.0cm,angle=0]{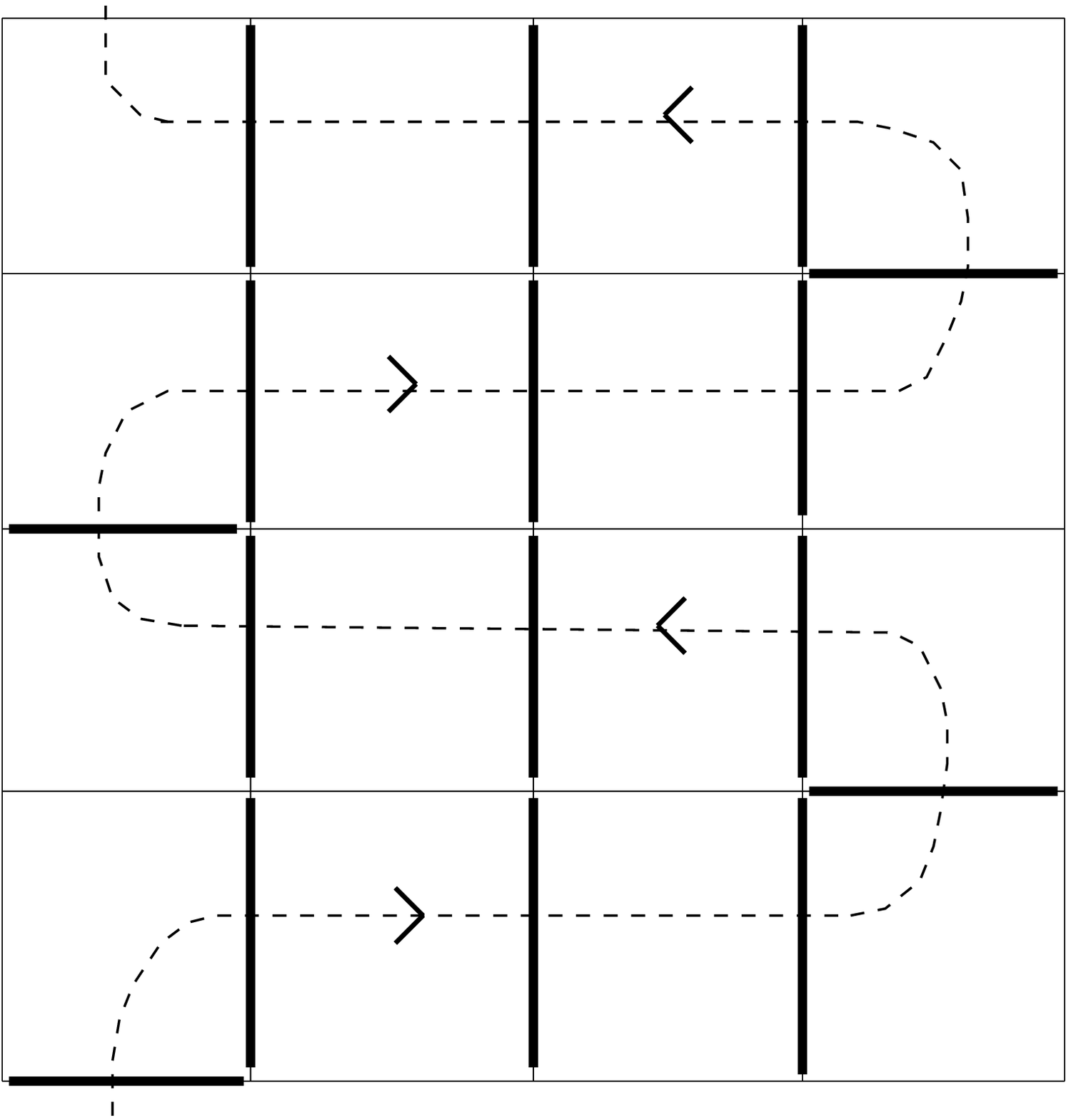}
\end{center}
\caption{\label{fig:snake_update} Picture of a plane through which
an extra $2\pi$ flux is imposed. The set of $L_\mu L_\nu$ links
that are modified is indicated with thick lines. The quantity
$\frac{2\pi}{L_\mu L_\nu}k, k=1,\ldots,L_\mu L_\nu$ is added to
the $k$-th link crossed by the dashed line.}
\end{figure}

Finally, in order to increase the flux tunneling probability, we
implemented a global update of the lattice in the following way
\cite{Heller}. Consider the $L_\rho L_\sigma$ planes in  the
orientation $(\mu,\nu)$, and for each plane select $L_\mu L_\nu$
links, as shown in Fig.~\ref{fig:snake_update}. Following the path
we draw, we add to the $k$-th link the quantity $\frac{2\pi}{L_\mu
L_\nu} k, k=1,\ldots,L_\mu L_\nu$, so that to each plaquette in
the plane an extra flux $\frac{2\pi}{L_\mu L_\nu}$ is imposed.
Overall, we add a flux $2\pi$ through the chosen orientation. To
satisfy detailed balance the orientation of the extra flux is
chosen randomly, and a Metropolis step is implemented. Although
the acceptance probability of such a global step is normally very
small, in our multicanonical approach it was greatly enhanced,
thus providing a valuable tool to decorrelate the flux faster.

\section{Twisted boundary conditions: U(1) vs. SU(N)}
\label{sec:Charact_of_phases}

We extend to the Abelian context the idea of 't Hooft
\cite{'tHooft:1979uj} about non-Abelian gauge theories: we probe
the response of the system to a variation of the flux. In the
context of  $SU(N)$ pure gauge theories, we know that it is
possible to modify the flux (by  twisting the boundary conditions)
according to the discrete group $Z_N$, center of $SU(N)$. This
modification arises from the requirement of changing the boundary
conditions of a gauge system defined on a hypertorus $T^4$ in the
most general way compatible with gauge invariance. We can
implement a similar change also in the Abelian context, but now
the difference is that we can vary the flux \emph{continuously}.
For this reason we will intentionally use the word `twist' as a
synonym of `flux'.\footnote{For an earlier, related study of compact
$U(1)$ in an external field, see \cite{Cea:1989ag}.}

\begin{figure}[h]
\begin{center}
\includegraphics[height=3.5cm]{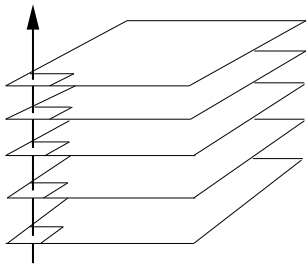}
\end{center}
\caption{\label{fig:twist_flux_picture}$3d$ representation of the
'twists' in $U(1)$ gauge theory. Each ($\mu,\nu$) plane now
carries flux $\phi$.}
\end{figure}

In order to implement this idea consider a stack of plaquettes,
one in each plane of a given orientation, as indicated (for the
$3d$ case) in Fig.~\ref{fig:twist_flux_picture}, and change their
value as:

\begin{equation}
\theta_P \to \theta_P+\phi, \quad \phi \in R
\end{equation}

\noindent
The new partition function is

\begin{equation}
\label{eq:twisted_partition_function} Z(\phi)=\int
D\theta\hspace{0.05cm} e^{\sum_{stack}(\beta\cos
(\theta_p+\phi)+\gamma \cos (2(\theta_p+\phi))+
\sum_{\overline{stack}}(\beta \cos \theta_p + \gamma\cos
2\theta_p)}
\end{equation}

\noindent
where \emph{stack}  indicates the set of plaquettes
through which the extra flux is imposed. For instance

\begin{equation}
stack=\lbrace  \theta_{\mu\nu}(x,y,z,t) \mid  \mu=1,\nu=2; x=1,y=1
\rbrace
\end{equation}

\begin{figure}[h]
\begin{center}
\begin{minipage}{5.5cm}
\begin{center}
\includegraphics[height=4.5cm,angle=0.]{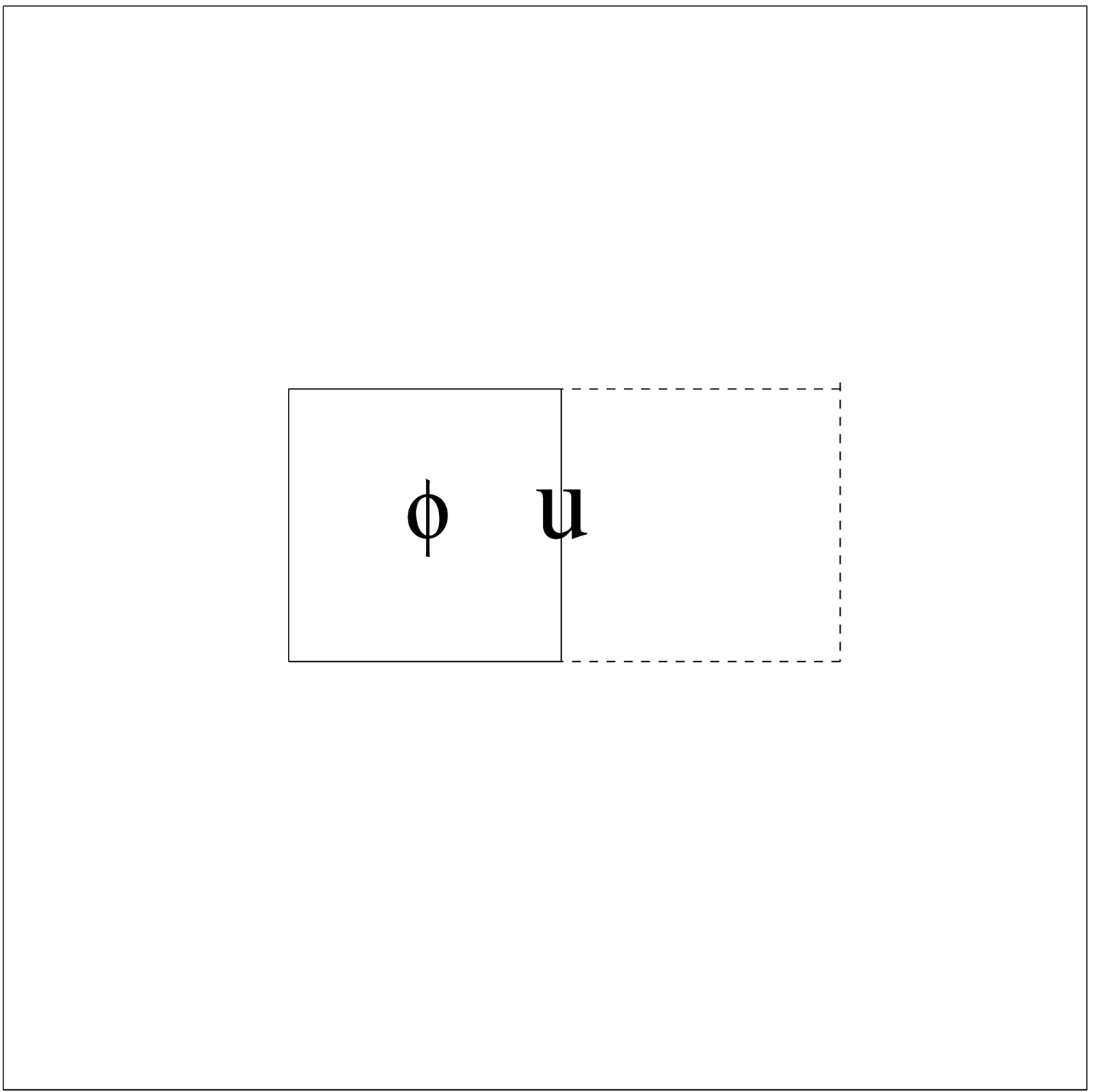}
\end{center}
\end{minipage}
\begin{minipage}{3.cm}
\begin{center}
$u \to u'=u \hspace{0.05cm} e^{-i\phi}$
\end{center}
\end{minipage}
\begin{minipage}{5.5cm}
\begin{center}
\includegraphics[height=4.5cm,angle=0.]{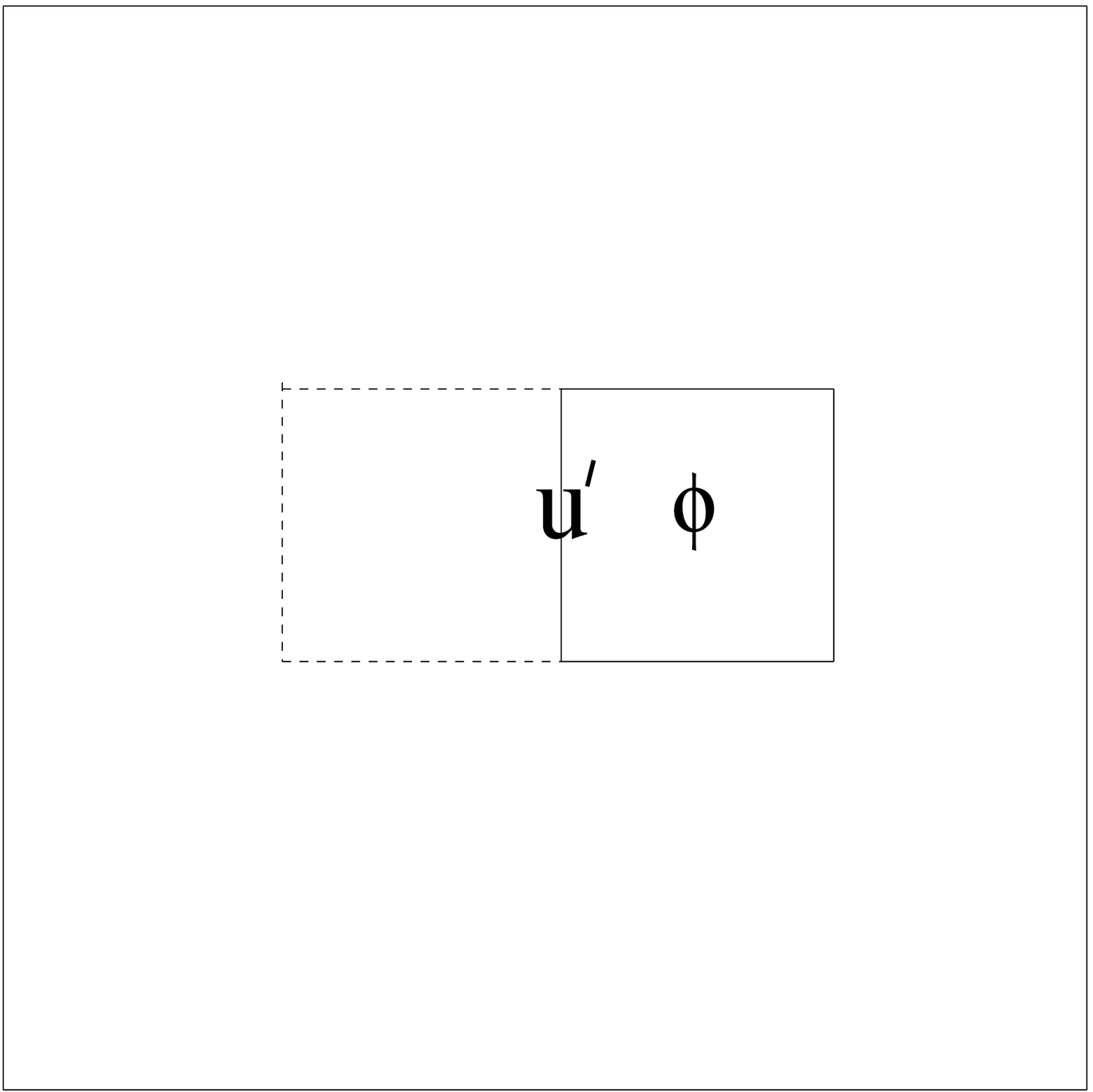}
\end{center}
\end{minipage}
\end{center}
\caption{\label{fig:tw_flux_1_2} Graphical representation of the change of variables that
moves the twisted stack. $u$ indicates a link; $\phi$ is the twist imposed on the
plaquette.}
\end{figure}

\noindent
'$\overline{stack}$' is its complement, that is all the
other unchanged plaquettes. Observe that $Z(\phi)$ is clearly
$2\pi$ periodic, so actually the flux we add to the system is
defined only $mod~ 2\pi$. It is not manifestly translationally
invariant, but it is easy to show that we can freely move the
position of the stack through a redefinition of the link
variables. In Fig.~\ref{fig:tw_flux_1_2} we show a two-dimensional
cartoon of this change of variables: $u$ is a link, and $\phi$
inside the plaquette indicates the extra flux. A change $u \to
e^{-i\phi}u$ moves the stack to a neighboring position.
\noindent
With a similar transformation (namely, moving only part
of the flux from the initial plaquette) one can spread the flux
arbitrarily over several plaquettes, in particular, homogeneously
through the whole plane.\\

Observe that we can implement the presence of an extra flux also
in terms of \emph{modified ("twisted") boundary conditions}

\begin{equation}
\label{eq:flux_and_boundary_cond}
U_2(x+L_1,y=1,z,t)=U_2(x,y=1,z,t)\hspace{0.05cm}e^{i\phi}
\end{equation}

\noindent
This remark will play an important role in the following.\\

We measure the free energy of the flux, defined by

\begin{equation}
\label{eq:flux_free_energy} F(\phi)=-\log \frac{Z(\phi)}{Z(0)}
\end{equation}

\noindent
and the results are as follows:

\begin{enumerate}

\item  In the confined phase (Fig.~\ref{fig:flux_FE_coul_conf},
lowest curve), within our statistical error, the free energy is
independent of $\phi$. The reason for this phenomenon is clear
from Eq.~(\ref{eq:flux_and_boundary_cond}). Since the flux
variation can be interpreted as a modification of the boundary
conditions, we expect that in the confined phase the system is
insensitive to them, because the correlation length $\xi$ is
finite (finite size corrections to $F(\phi)$ are ${\cal
O}(e^{-L/\xi}))$.

\item In the Coulomb (massless) phase the interaction is
long-range and the system becomes sensitive to the boundary
conditions (Fig.~\ref{fig:flux_FE_coul_conf}, upper curves).

It is possible to describe this behavior in terms of classical
arguments: we write the partition function of the system as a
\emph{sum over flux sectors} of the classical partition functions
corresponding to each sector, as derived in
Eq.~(\ref{eq:classical_flux_action})

\begin{equation}
\label{eq:classical_flux_FE_ansatz} Z(\phi)= \sum_k
Z_{cl.}(\phi+2\pi k)= \sum_k
e^{-\frac{\beta_{\rm{eff}}}{2}(\phi+2\pi k)^2}
\end{equation}

\noindent
We fit the flux free energy according to
Eq.~(\ref{eq:classical_flux_FE_ansatz}), letting
$\beta_{\rm{eff}}$ be a free parameter. The good description of
the data provided by this ansatz (see
Fig.~\ref{fig:flux_FE_coul_conf}) motivates the interpretation of
$\beta_{\rm{eff}}$ as an \emph{effective} coupling, which takes
into account the quantum fluctuations (that is, the contribution
of the monopole excitations) of the system. In section
\ref{sec:The helicity modulus} we will discuss this point in more
detail.

\begin{figure}[h]
\begin{center}
\includegraphics[height=10.0cm,angle=-90]{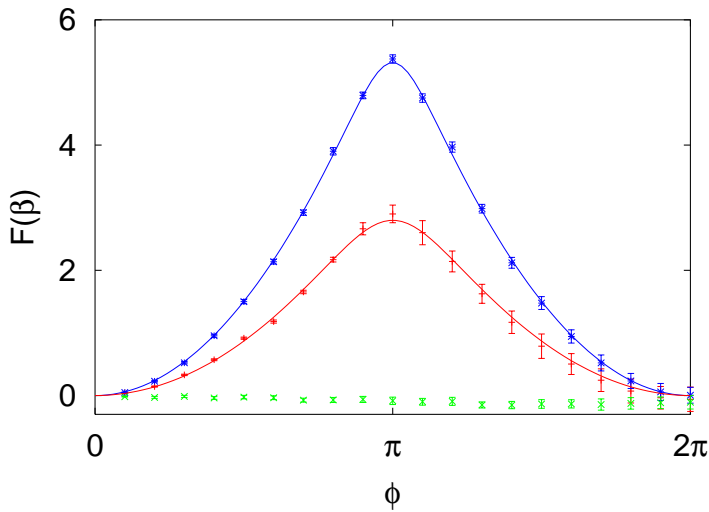}
\end{center}
\caption{\label{fig:flux_FE_coul_conf} Free energy $F(\phi)$ in
the Coulomb phase ($\beta=1.5,1.1$, upper and middle curve) and in
the confined phase ($\beta=0.8$, lower curve) ($4^4$ lattice). The
fit by the ansatz Eq.~(\ref{eq:classical_flux_FE_ansatz}) is
superimposed to the data. The effective couplings in the Coulomb
phase are $\beta_{\rm{eff.}}=1.2, 0.71$ respectively.}
\end{figure}

\item  In the unknown phases $II$ and $IV$ we observe (within the
statistical error) the appearance of an extra $\pi$ periodicity
(Fig.~\ref{fig:flux_FE_unknown_ph}). We now propose an argument to
interpret this finding.

\begin{figure}[h]
\begin{center}
\includegraphics[height=10cm,angle=-90]{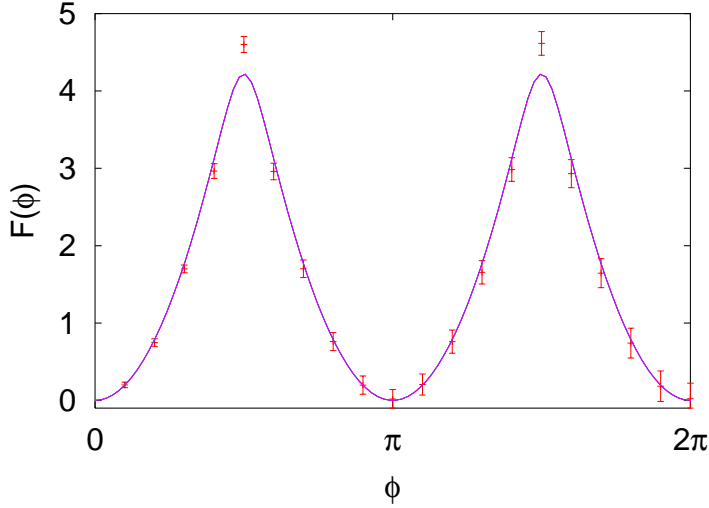}
\end{center}
\caption{\label{fig:flux_FE_unknown_ph} Typical free energy profile in phases $II$ and $IV$
($\beta=1.0$, $\gamma=-1.5$; $4^4$ lattice). The fit by the ansatz
Eq.~(\ref{eq:classical_flux_FE_ansatz}) (with the change $2\pi \rightarrow \pi$) is
superimposed (giving a value of $\beta_{\rm{eff}} \sim 3.98$).}
\end{figure}

\end{enumerate}


\noindent
Consider the following Polyakov loop correlator

\begin{equation}
\label{eq:poly_loop_correlator} \langle P_q(x)P_q^\dagger(x+L_\mu
\hat e_\mu)\rangle_{\tilde Z}
\end{equation}

\noindent
Here $P_q$ is the Polyakov loop of charge $q$ in
direction $\hat\nu$

\begin{equation}
P_q(x)=e^{iq \sum_{l=1}^{L} \theta_\nu(x+l\hat e_\nu)}
\end{equation}

\noindent
We take the average w.r.t. $\tilde Z=\int_0^{2\pi}d\phi
Z(\phi)$, in order to implement the influence of all possible
external fluxes ($mod ~2\pi$) on the system; the physics anyway
does not change. Even though the correlator is measured between
point $x$ and its periodic image $x+L_\mu\hat e_\mu$, it is
non-trivial due to the 'twisted' b.c.
(Eq.~(\ref{eq:flux_and_boundary_cond})). Gauss law gives

\begin{equation}
P_q(x+L_\mu \hat e_\mu)= e^{-iq\phi}P_q(x)
\end{equation}

\noindent
that is the two Polyakov loops differ by the flux we
added. Inserting this relation in
Eq.~(\ref{eq:poly_loop_correlator}) we get

\begin{equation}
\langle P_q(x)P_q^\dagger(x+L\hat e_\mu)\rangle_{\tilde Z}=
\frac{\int_0^{2\pi}d\phi \hspace{0.1cm}e^{iq\phi}
Z(\phi)}{\int_0^{2\pi}d\phi Z(\phi)}
\end{equation}

\noindent
If $Z(\phi)$ has periodicity $\pi$, it follows that when
$q=1$, or more generally odd, the numerator vanishes and the
correlator is zero, while if $q$ is even, it can be different from
zero. Noting that

\begin{equation}
\langle P_q(x) P_q^\dagger(x+L_\mu\hat e_\mu)\rangle\sim
e^{-V_q(L_\mu)L_\nu}
\end{equation}

\noindent
where $V_q$ is the static potential between the static
charges $\pm q$ represented by the Polyakov loops, a vanishing
Polyakov loop correlator implies an infinite (confining)
potential.

We therefore claim that the two unknown phases $II$ and $IV$ are
just Coulomb phases for \emph{even} electric charges, while odd
charges are confined and paired in even combinations.

Maybe a simpler argument can start from the observation that this result is trivially true
in the Coulomb phase along the axis $\beta=0$, where the $\pi$ periodicity is a simple
consequence of the symmetry $(\beta,0)\to(0,\beta)$ discussed in Sec.~\ref{sec:the_model}.
It then easily extends to the whole regions $II$ and $IV$ of Fig.\ref{fig:phase_diagram},
since a phase is characterized by a given infra-red (IR) behavior. Indeed our numerical
results (obtained at $\beta=1.0$ and $\gamma=-1.5$) (Fig.~\ref{fig:flux_FE_unknown_ph})
show no detectable deviation from $\pi$ periodicity.

The extra periodicity is also a useful tool to discuss the fate of
the two phase boundaries at the lower corners of
Fig.~\ref{fig:phase_diagram}. It is possible that these phase
boundaries meet at some point or that they just become
asymptotically close to each other. What we can exclude is that
they meet and terminate somewhere, letting the two different Coulomb
phases communicate. Suppose that they communicate somewhere, then
we expect the twisted free energy $F(\phi)$
(Fig.~\ref{fig:flux_FE_coul_conf} and
\ref{fig:flux_FE_unknown_ph}) to be the same in both: but this
leads to a contradiction. In fact $F(\phi)$ would have
\emph{least} periodicity $\pi$ and $2\pi$ at the same time. This
is only possible if the distribution is flat, which only happens
in the confined,
not the Coulomb, phase.\\

We conclude this section with some comments about the numerical
strategy used to compute the free energy
Eq.~(\ref{eq:flux_free_energy}). Here we are faced with an
\emph{overlap} problem, that is, in the ratio
$\frac{Z(\phi)}{Z(0)}$, the numerator and the denominator sample
non-overlapping regions of the configuration space. We can cure
this situation through the following strategy
\cite{deForcrand:2000fi}. Divide the interval $[0,2\pi)$ into $N$
parts,  and consider the identity

\begin{equation}
\label{eq:chain_rule}
\frac{Z(2\pi)}{Z(0)}=\frac{Z_N}{Z_{N-1}}\frac{Z_{N-1}}{Z_{N-2}}
\ldots\frac{Z_1}{Z_0}
\end{equation}

\noindent
where $Z_k$ is defined as the partition function of a
system with an extra flux $\phi~=\frac{2\pi}{N}k$.  We observe
that for each ratio the overlap between numerator and denominator
is much better now. A further improvement of the overlap (and,
therefore, further variance-reduction) is obtained if we express
each ratio as

\begin{equation}
\label{eq:second_ratio}
\frac{Z_k}{Z_{k-1}}=\frac{Z_k/Z_{k-\frac{1}{2}}}{Z_{k-1}/Z_{k-\frac{1}{2}}}
\end{equation}

\noindent
where the index $k-\frac{1}{2}$  refers to an external
flux $\phi=(k-\frac{1}{2})\hspace{0.1cm}\frac{2\pi}{N}$. The
numerator and the denominator of each ratio are ratios themselves,
where, for example,

\begin{equation}
\frac{Z_k}{Z_{k-\frac{1}{2}}}=\langle
e^{-(S_k-S_{k-\frac{1}{2}})}\rangle_{Z_{k-\frac{1}{2}}}
\end{equation}

\noindent
$S_k$ is the action of the system when an external flux
$k\hspace{0.1cm}\frac{2\pi}{N}$ is added. This  ratio can be
directly computed. The algorithm consists of measuring all these
ratios separately and recombining them according to
Eqs.~(\ref{eq:chain_rule})(\ref{eq:second_ratio}). It is a trivial
matter to extract statistical errors since the $N$ ratios are
estimated by independent Monte Carlo simulations.


\section{The helicity modulus}
\label{sec:The helicity modulus}

We now go back to Fig.~\ref{fig:flux_FE_coul_conf}, i.e., to the
free energy of a system with twist $\phi$. Consider the
\emph{curvature} of the free energy profile at some arbitrary
value of the flux $\phi=\bar\phi$. In the confined phase it is
identically zero (in the thermodynamic limit), because of the
independence of $F(\phi)$ from $\phi$. In the Coulomb phase this
quantity is different from zero. Therefore there is certainly a
point of non-analyticity for this quantity at the transition: it
is an \emph{order parameter}.

\noindent
We are then motivated to introduce the following
definition

\begin{equation}
\label{eq:helicity_modulus} h(\beta,\gamma)=\frac{\partial^2
F(\phi)}{\partial \phi^2} \mid_{\phi=\bar\phi=0}
\end{equation}

\noindent
known as the \emph{helicity modulus} in the literature \cite{Jose:1977gm}
\cite{Cardy:jg} (we stress that the choice $\bar\phi=0$ is arbitrary). This name was
conceived in the context of the XY (rotor) model, where the boundary conditions are changed
by adding an angle $\phi$ to the rotor phase $\theta(x,y)$ \cite{Cardy:jg}

\begin{equation}
\label{eq:rotor_twist} \theta(x+L_1,y)=\theta(x,y)+\phi
\end{equation}

\noindent
so that the quantity (\ref{eq:helicity_modulus}) assumes
the meaning of the response function (`modulus') to a rotation
(`helicity') of the boundaries. Eq.~(\ref{eq:rotor_twist}) is the
analog of Eq.~(\ref{eq:flux_and_boundary_cond}) in our case.

\noindent
If we substitute our partition function $Z(\phi)$
Eq.~(\ref{eq:twisted_partition_function}) into
Eq.~(\ref{eq:helicity_modulus}) we obtain the following expression

\begin{equation}
\label{eq:helicity_modulus_explicit} h(\beta,\gamma)=\langle
\sum_{stack}(\beta\cos[\theta_P]+4\gamma\cos[2\theta_P])\rangle
-\langle(\sum_{stack}(\beta\sin[\theta_P]+2\gamma\sin[2\theta_P]))^2\rangle
\end{equation}

\noindent
where `\emph{stack}' refers to the stack of plaquettes
we changed by adding the flux (see Section
\ref{sec:Charact_of_phases}). As we noted earlier, we are allowed
to spread the flux homogeneously through the planes via a
coordinate transformation. If we add to each plaquette of a given
orientation the flux $\phi/L_\mu L_\nu$ and recompute
Eq.~(\ref{eq:helicity_modulus}), we obtain equivalently

\begin{eqnarray}
\label{eq:helicity_modulus_spread}
h(\beta,\gamma)&=&\frac{1}{(L_\mu L_\nu)^2}\lbrace\langle
\sum_{P}(\beta\cos[\theta_P]+4\gamma\cos[2\theta_P])\rangle
\nonumber \\
&-&\langle(\sum_{P}(\beta\sin[\theta_P]+2\gamma\sin[2\theta_P]))^2\rangle\rbrace
\end{eqnarray}

\noindent
where now the sum extends over all plaquettes in the
$(\mu,\nu)$ orientation. This expression turned out to give a much
smaller variance than the other one.

Suppose that $\gamma=0$. Then it is interesting to observe that in
the confined phase, where this order parameter is zero, the
average action (first term) is related to a measure of its
fluctuations (second term).

\section{Helicity modulus: numerical results}
\label{sec:Helicity_modulus_numerical}

\begin{figure}[h]
\begin{center}
\includegraphics[height=10.0cm,angle=-90]{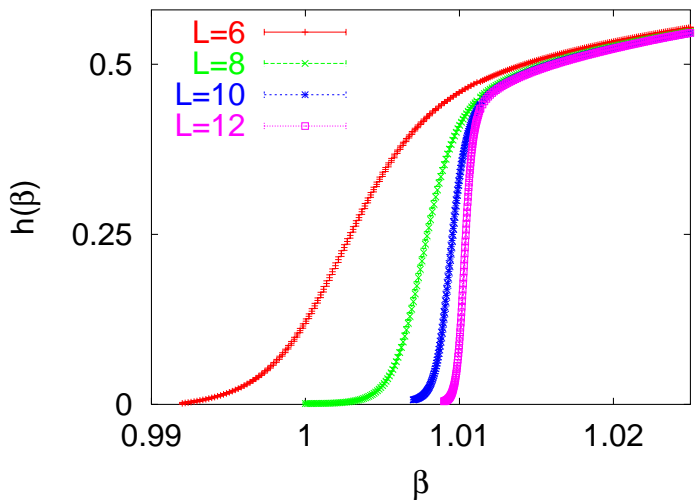}
\end{center}
\caption{\label{fig:helicity_modulus_plot} Helicity modulus on the
Wilson axis near the phase transition.}
\end{figure}

We now present the results of a numerical study of the helicity
modulus. First, in Fig.~\ref{fig:helicity_modulus_plot}, we show
the behavior of this quantity on the Wilson axis. Remaining for
the moment on a qualitative footing, we see that it indeed behaves
like an order parameter. On the extended line ($\beta=0$) we would
obtain the same plot up to a factor $4$, as is clear looking at
Eq.~(\ref{eq:helicity_modulus_explicit}) and remembering the
symmetry $(\beta,0)\leftrightarrow (0,\beta)$ of the partition
function $Z(\beta,\gamma)$.

The statistics used for this measurement is of about $10^5$
measurements per lattice size, with one heatbath and four
overrelaxation sweeps between two subsequent measurements. The
curves in Fig.~\ref{fig:helicity_modulus_plot} are obtained
through a reweighting of the data $\grave{a}$ $la$
Ferrenberg-Swendsen
\cite{Ferrenberg:yz}.\\

\begin{figure}[h]
\begin{center}
\includegraphics[height=10.0cm,angle=-90]{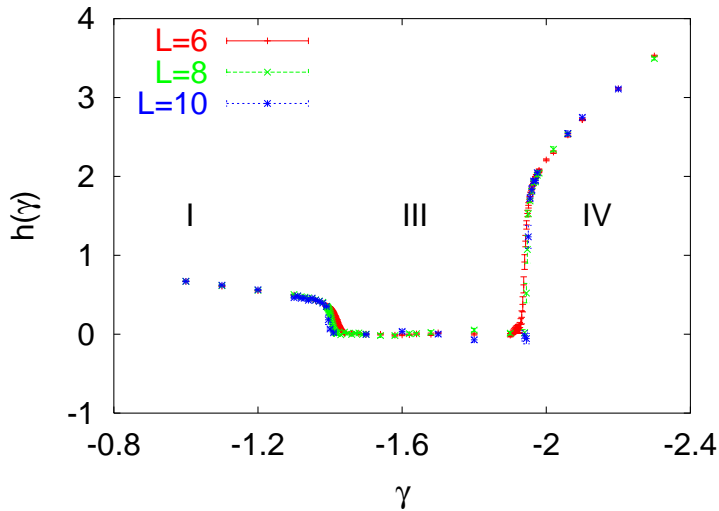}
\end{center}
\caption{\label{fig:hm_ext_plane}Helicity modulus versus $\gamma$
($\beta=2.5$ fixed; $L=6,8,10$). Label $`I'$ indicates the usual Coulomb
phase, $`III'$ the confined phase, $`IV'$ the even-charge Coulomb phase.}
\end{figure}

Let us now consider the behavior of the helicity modulus in the
whole phase diagram. Again refer to Fig.~\ref{fig:phase_diagram},
fix the value of $\beta=2.50$, for example, and consider the
(negative) values of $\gamma$ in a region across the phase
boundaries of the diagram. We  show the result in
Fig.~\ref{fig:hm_ext_plane}. Also in this case the expectations
are qualitatively satisfied: the flat region signals the presence
of a confining phase, and the helicity modulus is roughly $4$
times higher in the even-charge Coulomb phase than in the other
one.

From the numerical point of view, even if the statistics used is
roughly the same as on the Wilson axis (at least across the phase
transitions), these data are less accurate than the other ones: at
such a small coupling the Monte Carlo dynamics is slower (much
larger autocorrelation time).

In the next sub-sections we perform a quantitative analysis of these data, in order to
clarify the issues of the order of the phase transition, of the relation between the
helicity modulus and the renormalized coupling, and finally of the conjectured universality
of the helicity modulus at $\beta_c$ \cite{Cardy:jg}.

\subsection{Order of the phase transition}
\label{sec:order_of_the_phase}

The difficulty of determining the order of a phase transition
using lattice simulations is well known, and, in general terms, is
due to the fact that in finite systems no singularities of the
free energy can be present. In our specific case we are interested
in distinguishing a first- from a second-order transition, but we
know that even in the first-order scenario the correlation length
$\xi_c$ at the transition point can be quite large, so that for
lattices of size $L \lesssim \xi_c$ the transition mimics a
second-order one. This can explain the difficulty that was
encountered in solving this problem so far.

As shown in Fig.~\ref{fig:helicity_modulus_plot} the finite-size
(FS) effects act to round off the singularity into smooth curves.
It turns out that their functional form depends on the
\emph{order} of the phase transition, so that an ansatz for the FS
effects, when available, provides a direct way to prove or
disprove a conjecture about the nature of the phase transition.
Since we have accurate data for the helicity modulus for lattices
of size up to $L=14$, we try to make such an analysis both for the
first- and the second-order scenarios.

From a theoretical point of view, little is known about the order
of the transition. If the correlation length would diverge at the
transition point (as happens in the second-order scenario), it
would be possible to extract a continuum limit, leading to a
continuum \emph{confining Abelian} theory, which seems unlikely.
From a numerical point of view, a quite recent study
\cite{Arnold:2002jk} of the $4d$ Abelian Wilson action strongly
supports the first-order scenario, through a high statistics
analysis of the action density distribution (considering lattices
of size up to $L=18$).\\

Let us first consider the first-order scenario, and 
make the hypothesis that the helicity modulus has a
\emph{jump} across the transition point. We now consider the
simplest ansatz for the partition function in the vicinity of the
transition point

\begin{equation}
\label{eq:Borgs_kotecky_ansatz} Z \sim
e^{-Vf_-(\beta)}+Xe^{-Vf_+(\beta)}
\end{equation}

\noindent
where $f_-,f_+$ denote the free energy densities of the two coexisting phases
(such that $f_-(\beta_c)=f_+(\beta_c)$), and $X$ is the so-called \emph{phase asymmetry
parameter}, that is the relative weight of the two phases at the coexisting point.
Eq.~(\ref{eq:Borgs_kotecky_ansatz}) is  the Borgs-Kotecky ansatz
\cite{Borgs:1990cz}\cite{Borgs:xs}, but we have to observe that the work of these two
authors is based on a Peierls-like contour picture of the system at the transition point,
with the hypothesis that the contours (that is, the boundary of domains characterized by a
single phase) do not interact with each other. The long-range nature of the photon
interaction in the Coulomb phase makes it difficult to justify such a hypothesis. A
posteriori, we will indeed observe deviations from the predictions based on this ansatz:
for example, finite-size corrections to the fitting parameters different from the standard
ones proportional
to $\frac{1}{V}$ will be necessary.\\

The ansatz of Eq.~(\ref{eq:Borgs_kotecky_ansatz}) leads to a
formula for the interpolating FS curves which we write in the
following way

\begin{equation}
\label{eq:hm_first_order_ansatz}
h(\beta,V)=\frac{h_+}{1+X^{-1}e^{-V\Delta f'\cdot
(\beta-\beta_c)}}
\end{equation}

\noindent
where $h_+$ is the value of the helicity modulus in the
pure Coulomb phase (see Fig.~\ref{fig:three_values_at_jump}), and
$\Delta f'=f'_+-f'_-$ is the \emph{latent heat}, obtained from the
expansion

\begin{equation}
\label{eq:expansion_free_energy}
f_\pm(\beta)=f_c+f'_\pm\cdot(\beta-\beta_c)+{\cal{O}}(\Delta
\beta^2).
\end{equation}

Since for simplicity $h_+$ is taken as constant in the Coulomb
phase, the purpose of the ansatz is to describe the helicity
modulus just below the transition, and we perform a fit for values
of $\beta \leq \beta_c$ (actually the range was limited to an
interval $[\beta_1,\beta_c]$, with $\beta_1$ close to $\beta_c$,
to take into account the approximation introduced by linearizing
Eq.~(\ref{eq:expansion_free_energy})). For $\beta_c$ we use the
very precise value in \cite{Arnold:2002jk}, and also we determine
it self-consistently from the fit of our data (iteratively, we fit
the data in an extended $\beta$ range, determine $\beta_c$ and
make another fit in a range up to the new value; we repeat this
procedure until $\beta_c$ stabilizes). An analysis of the possible
FS corrections indicates that a minimal choice is the following

\begin{eqnarray}
\label{eq:corrections_hm_ansatz_1}
h_+      &\to&   h_+        \cdot       (1 +  \frac{\alpha_1}{L})  \\
\label{eq:corrections_hm_ansatz_2} \Delta f' &\to&   \Delta f'
\cdot       (1 + \frac{\alpha_2}{V})
\end{eqnarray}

\begin{figure}[h]
\begin{center}
\includegraphics[height=9.0cm,angle=-90]{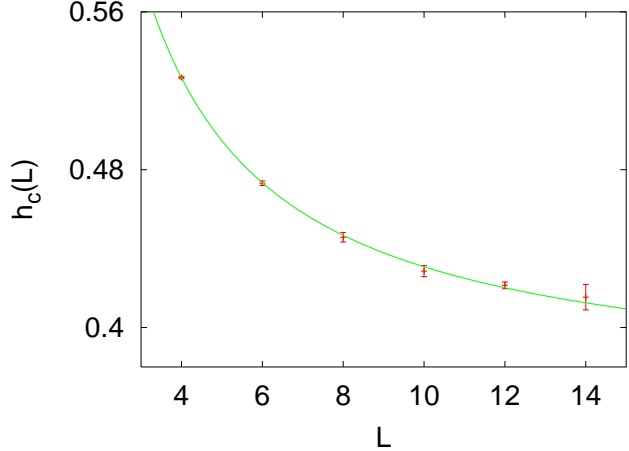}
\end{center}
\caption{\label{fig:hm_at_criticality} Helicity modulus at the
transition point versus lattice size.}
\end{figure}

\begin{figure}[t]
\begin{center}
\includegraphics[height=4.0cm,angle=0]{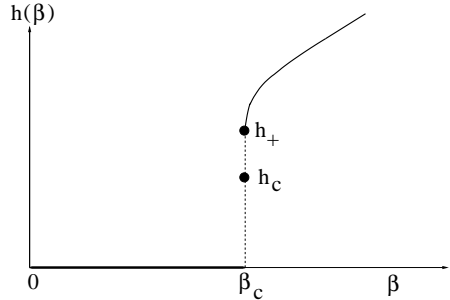}
\end{center}
\caption{\label{fig:three_values_at_jump} Graphical explanation of
the physical meaning of the parameters $h_+$ and $h_c\equiv
h(\beta_c,V=\infty)$, as defined in
Eq.~(\ref{eq:hm_first_order_ansatz}).}
\end{figure}

\noindent
For the \emph{latent heat} we consider $1/V$ corrections, standard for a
first-order transition. To justify the $1/L$ corrections used for $h_+$, first let us look
at the behavior of the helicity modulus at the transition point
(Fig.~\ref{fig:hm_at_criticality}).

We fit these data with the ansatz

\begin{equation}
h_c(L)=h_c(\infty)\cdot(1+\frac{c}{L^\nu})
\end{equation}

\noindent where $c$ is a constant, and obtain the result $\nu=1.04(12)$ , which is a
first numerical indication for the choice of the FS correction
Eq.~(\ref{eq:corrections_hm_ansatz_1}). At the end of this paragraph we are going to
comment more extensively about this choice. Then, fixing $\nu=1$, we get the thermodynamic
value $h_c(\infty)=\frac{h_+}{1+X^{-1}}=0.371(1)$ ($c=1.7$), independently from the
Borgs-Kotecky ansatz. This value is only slightly increasing if we
exclude from the fit the points corresponding to $L=4, 6$.\\

We then turn to the ansatz Eq.~(\ref{eq:hm_first_order_ansatz})
and fit  the data corresponding to lattices $L=8,10,12,14$,
obtaining the following results:

\begin{eqnarray}
\label{eq:fit_results}
h_+               &=& 0.381(5)     \\
\Delta f'/6       &=& 0.029(2)     \\
\label{eq:phase_asymm_par}
\log X            &=& 3.20(15)     \\
\beta_c           &=& 1.01108(5)   \\
\alpha_1          & \simeq & 1.8     \\
\alpha_2          & \simeq & (3.7)^4
\end{eqnarray}

\noindent
Comparing the quantities $\beta_c$, $\log X$ and $\Delta
f'/6$ (the latent heat per plaquette) with earlier measurements by
Arnold et al. \cite{Arnold:2000hf}\cite{Arnold:2002jk}, the
agreement is very good for all the values.

As a crosscheck, we can relate the value of $h_c(\infty)=0.371(1)$
measured before, and $h_+=0.381(5)$ in Eq.~(\ref{eq:fit_results}).
Their physical meaning is explained in
Fig.~\ref{fig:three_values_at_jump}: $h_+$ is defined as

\begin{equation}
h_+=\lim_{\beta\to\beta^+}h(\beta)
\end{equation}

\noindent
and its value depends on the Coulomb phase only, while
$h_c$ is smaller, and takes into account the coexisting confined
phase. As one can easily check, the relation

\begin{equation}
h(\beta_c)=\frac{h_+}{1+X^{-1}}
\end{equation}

\noindent
(Eq.~(\ref{eq:hm_first_order_ansatz}) computed at $\beta=\beta_c$) is consistent
with our numerical values.

We also measured the value of $h_+$ directly, extracting from the 
canonical ensemble at
$\beta_c$ the subset of configurations whose action density is lower than a threshold value
($0.359$) chosen to best separate the two phases; in this way one can probe the Coulomb
phase contribution only. We find $h_+=0.4325(15)$, a value incompatible with the value 
fitted using the BK ansatz. The discussion of this problem is postponed to the end of this 
section.\\

\begin{figure}[h]
\centerline{
\includegraphics[height=8.2cm,angle=-90.]{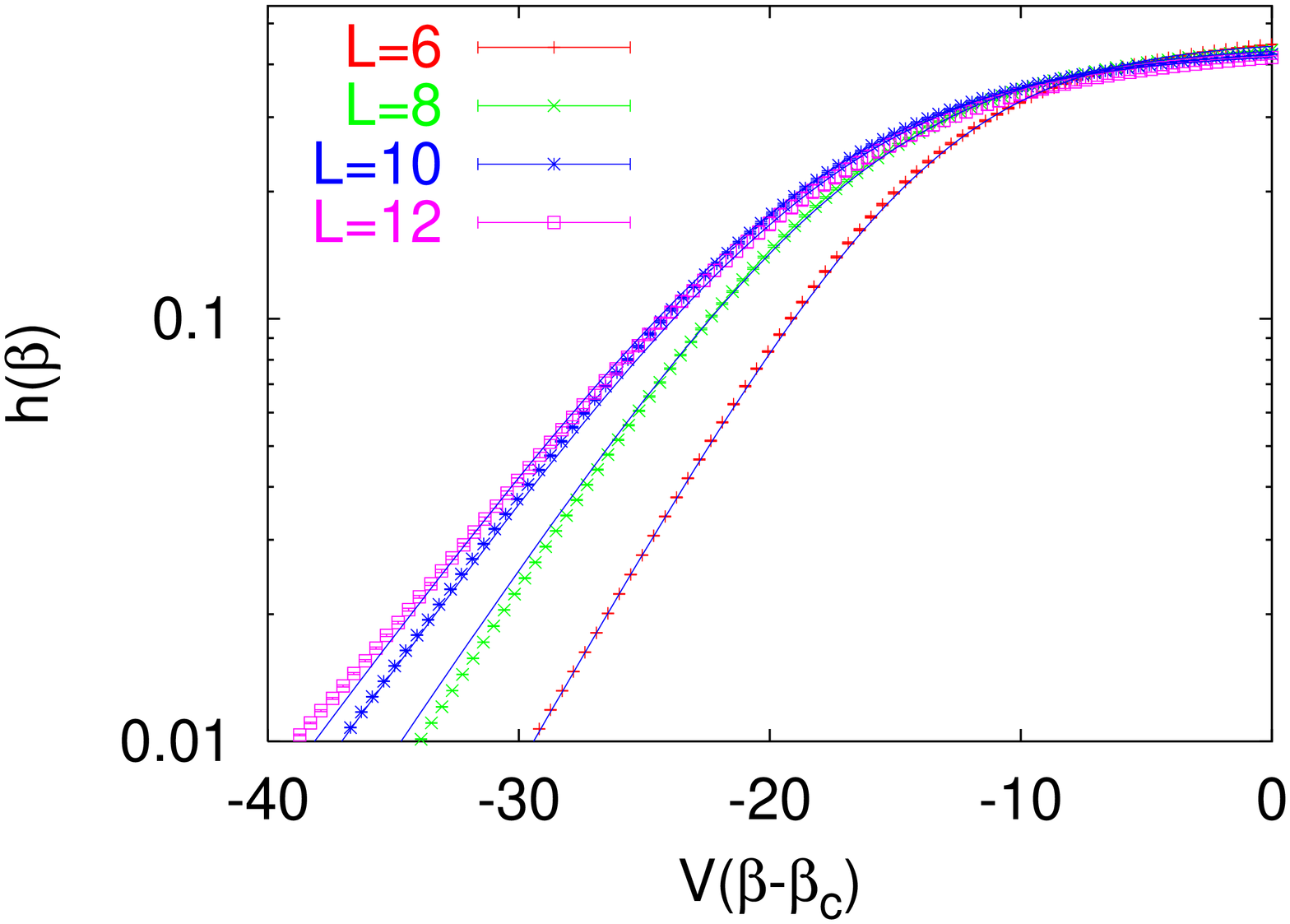}
\includegraphics[height=8.2cm,angle=-90.]{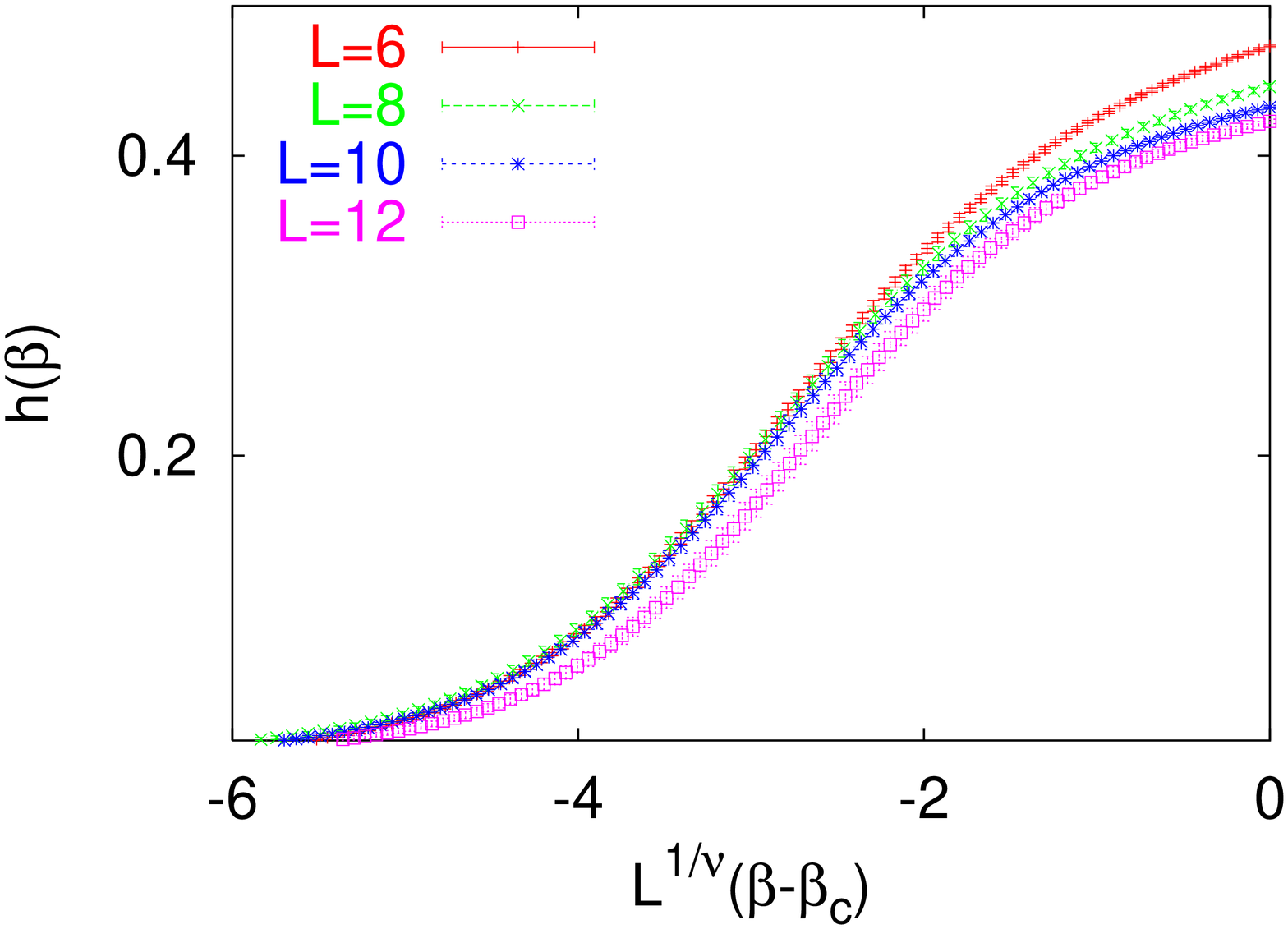}
}
\caption{\label{fig:hm_rew_fit_quality} Graphical comparison of
the quality of the first and second order ansatz for the helicity
modulus data. The solid lines (left) show the first-order fit
Eq.(\ref{eq:hm_first_order_ansatz}).}
\end{figure}

All the errors are computed performing a jackknife analysis. The
values of $\alpha_1, \alpha_2$ are mentioned, to give an idea of
the magnitude of the correction involved. Since we want to show
the relative correction that they induce, $\alpha_2$ (which has
dimension $L^4$) is expressed as a fourth power. To evaluate the
quality of the fit we could not use just the reduced $\chi^2$,
because the data we fit are re-weighted, so that the number of
points in the fit and their mutual correlation play a role which
is not properly taken into account by the value of $\chi^2$. So we
decided to show directly a plot (Fig.\ref{fig:hm_rew_fit_quality},
left) of the fitted data. The most important signature of the
first order nature of the transition is the exponential (and not
power law) behavior of the helicity modulus as a function of the
distance from the transition point, which is visible in the linear
behavior (in log-scale) away from
$\beta_c$ (see Fig.~\ref{fig:hm_rew_fit_quality}). \\

Now, for completeness, we want to compare the quality of this
ansatz with that of the second-order one. If the  transition would be
second-order, all quantities (in the vicinity of the critical
point) should satisfy the scaling hypothesis, that is they should
be a function of $L/\xi$ ($\xi$ is the correlation length) only.
Since

\begin{equation}
\xi \sim (\beta-\beta_c)^{-\nu}
\end{equation}

\noindent
a rescaling of $\beta$ according to

\begin{equation}
\label{eq:rescale_beta_second_order} \beta \to
L^{1/\nu}(\beta-\beta_c)
\end{equation}

\noindent
should produce a superposition (`collapse') of the data.
The result of this analysis is shown in
Fig.~\ref{fig:hm_rew_fit_quality} on the right for the `test'
value of $\nu=0.3$. The data collapse is quite poor: we consider
this as an indication that the second-order scenario can be
excluded.\\

We conclude this section with a discussion of the inconsistency we found between
the determinations of $h_+$ using the BK ansatz ($h_+ = 0.381(5)$) 
and the direct measurement ($h_+ = 0.4325(15)$).
On one hand, both methods are affected by systematic errors. Concerning the direct
measurement, notice that a single metastable phase is not really
a pure phase, but contains small `bubbles' of the other one; furthermore the method is
sensitive to the threshold value, which is chosen somewhat arbitrarily between the two
peaks of the action density distribution. However, these two kinds of error diminish
as the thermodynamic limit is approached; thus we were led to complement the other 
simulations with one on a large lattice ($32^4$) in order to consider  these errors as small. 
Let us now scrutinize the BK ansatz.
First of all we can consider the possibility that the lattices we used are too small, so
that significant higher order $1/L^k, k>1$ corrections should be added to the leading
terms; but if we exclude from the fit the smallest lattices ($6^4,8^4,10^4$), no clear
increase in the fitted value of $h_+$ is observed. Similarly, no clear change in the value
of the fitting parameters occurs if we include a FS correction for $X$, the phase asymmetry
parameter, not considered in
Eqs.~(\ref{eq:corrections_hm_ansatz_1})(\ref{eq:corrections_hm_ansatz_2}). It appears that
the BK ansatz does not describe so well the data very near the transition. The reason
should be found, as we already mentioned, in the long-range nature of the photon
interaction, which is not taken into account by this ansatz based on a strong first-order
picture. Note that corrections of order up to $1/V^4$ were found necessary in
\cite{Arnold:2002jk} to describe the behavior of the pseudocritical $\beta_c(L)$ on
lattices of similar sizes to ours. From this discussion we also obtain some plausible motivation 
for the unusual FS correction ($1/L$ instead of $1/V$ at the leading order 
(Eq.(\ref{eq:corrections_hm_ansatz_1}))) we observe for the helicity modulus at the transition; 
we can impute such large correction to the interacting interfaces between the coexisting phases. 

Therefore, we blame our inconsistency between the fitted
and the measured value of $h_+$ on the limitations of the BK ansatz. The direct measurement of 
$h_+$ is independent of any specific model of the system at the transition, 
and the systematic  errors 
that affect its value are certainly decreasing as the thermodynamic limit is approached. 
Thus, whenever possible, we will use this determination.
 



\subsection{The helicity modulus and the renormalized coupling}
\label{sec:the _rinormalized_coupling}

We now want to discuss a remarkable property of our order parameter \cite{Cardy:jg}: the
helicity modulus in the Coulomb phase is related to the value of the renormalized coupling.
Let us go back to the classical Eq.~(\ref{eq:action_quadr_in_flux})

\begin{equation}
\label{eq:integrated_hm}
F(\phi)-F(0)=\frac{\beta}{2}\phi^2
\frac{L_\rho L_\sigma}{L_\mu L_\nu}
\end{equation}

\noindent
where, as already defined, $F(\phi)$ is the flux free energy. Following
\cite{Cardy:jg} we can define a so-called `blocking transformation'. As the number of RG
steps increases and short distances are integrated out, only one relevant, Coulombic
coupling survives, while all other couplings go to zero. In this way, starting from
Eq.~(\ref{eq:integrated_hm}), we end up with the result

\begin{equation}
\label{eq:renorm_near_to_cl_limit}
F(\phi)-F(0)=\frac{\beta^R}{2}\phi^2\frac{L_\rho L_\sigma}{L_\mu
L_\nu}
\end{equation}

\noindent
where the bare coupling has been substituted by the
renormalized one, but no other changes occurred. The second
derivative of this quantity thus gives the renormalized coupling,
as anticipated.

\begin{figure}[h]
\begin{center}
\includegraphics[height=8.0cm,angle=-90]{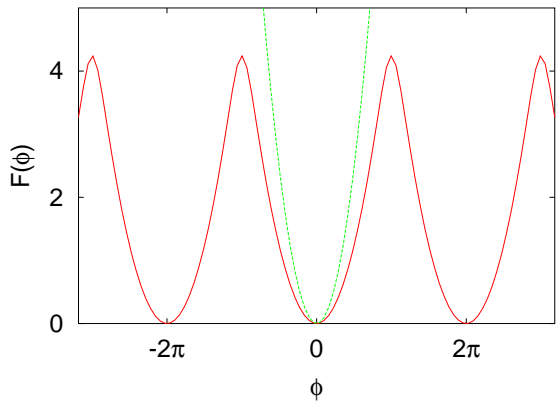}
\end{center}
\caption{\label{fig:flux_sector_correction} Qualitative behavior
of the ansatz Eq.~(\ref{eq:free_energy_for_fluxes}) and
Eq.~(\ref{eq:renorm_near_to_cl_limit}); the difference disappears
in the limit $\beta \to \infty$.}
\end{figure}

The next step is to observe that the functional form of
Eq.~(\ref{eq:renorm_near_to_cl_limit}) is not totally consistent:
this equation does not present any periodicity in $\phi$, but we
know that the flux free energy is $2\pi$ periodic in $\phi$, the
external flux we add, as already visible in
Eq.~(\ref{eq:twisted_partition_function}). Therefore we have to
consider the influence of all the other flux sectors, according to
Eq.~(\ref{eq:classical_flux_FE_ansatz}), and write

\begin{equation}
\label{eq:free_energy_for_fluxes} F(\phi)= -\log(\sum_k
e^{-\frac{\beta_{R}}{2} \cdot (\phi+2\pi k)^2})
\end{equation}

\noindent
which correctly displays the periodic structure of the
flux free energy. In Fig.~\ref{fig:flux_sector_correction} we show
a cartoon comparison of equations
(\ref{eq:renorm_near_to_cl_limit}) and
(\ref{eq:free_energy_for_fluxes}).

\noindent
Therefore we define $\beta_R(\beta)$ implicitly through
the equation

\begin{equation}
\label{eq:hm_and_renorm_coupling} \frac{\partial^2 F(\phi,\beta_R)}{\partial
\phi^2}\mid_{\phi=0}=h_0(\beta)\equiv h(\beta)
\end{equation}

\noindent
where $F(\phi,\beta)$ is given in Eq.~(\ref{eq:free_energy_for_fluxes}). In order
to give an idea of how the map $\beta_R(\beta)$ works, observe that in the limit
$\beta\to\infty$ the following result holds

\begin{equation}
\beta_R(\beta)=\beta[1-\frac{1}{4\beta}+{\cal{O}}(\frac{1}{\beta^2})]
\end{equation}

\noindent
because as $\beta\to \infty$ the renormalized coupling is equal to the helicity
modulus up to exponentially small corrections, hence we can use the weak coupling expansion of
Sec.\ref{sec:universality_of_the_hm}.

In section  \ref{sec:Charact_of_phases} we already showed that the
ansatz Eq.~(\ref{eq:free_energy_for_fluxes}) provides an accurate
description of the numerical data. Here we make a more refined
check: this ansatz has to provide the right mapping between the
helicity modulus at flux $\phi=0$ and $\phi=\pi$. Since it depends
on only one parameter, once the curvature at the origin is fixed,
the curvature at $\phi=\pi$ has to be fixed too. In Table
\ref{tab:hm_at_flux_zero_and_pi} we check the validity of this
assumption. As we can see, the ansatz works very well over a wide
range of $\beta$ values. We also observe that from the theoretical
point of view, defining the helicity modulus as the curvature of
the flux free energy at flux $\pi$, instead of at flux $0$, could
have been a perfectly consistent choice:

\begin{equation}
\label{eq:hm_and_renorm_coupling_at_flux_pi} \frac{\partial^2 F(\phi,\beta_R)}{\partial
\phi^2}\mid_{\phi=\pi}=h_\pi(\beta)
\end{equation}

\noindent
Therefore,
Eq.(\ref{eq:hm_and_renorm_coupling_at_flux_pi}) provides an
equivalent way to compute the renormalized  coupling.
Moreover, the data obtained at flux $\pi$ are statistically more accurate.\\

\begin{table}
\begin{center}
\begin{tabular}{|c|c|c|c|}\hline

$\beta$ & $h_{\phi=0}(\beta) $ &
$\frac{d^2 F}{d\phi^2}(\beta)\mid_{\phi=\pi}$ & $h_{\phi=\pi}(\beta)$  \\
    &  & \small{\rm{derived from col.2}} & \small{\rm{measured}} \\

\hline

1.02 & 0.524(1) & -2.18(1) & -2.16(1)   \\ \hline

1.10 & 0.713(1) & -4.30(1) & -4.30(1)   \\ \hline

1.20 & 0.852(1) & -6.31(1) & -6.32(1)   \\ \hline

1.30 & 0.972(1) & -8.35(2) & -8.32(1)   \\ \hline

\end{tabular}
\caption{\label{tab:hm_at_flux_zero_and_pi} Check of the validity
of the  ansatz Eq.~(\ref{eq:free_energy_for_fluxes}). The first
column is the bare (inverse) coupling; the second column is the
measured value of the helicity modulus at flux $0$; the third and the fourth columns
correspond, respectively, to the analytic prediction  of the
helicity modulus at flux $\pi$, derived from that at flux $0$
according to Eq.~(\ref{eq:hm_and_renorm_coupling}), and to the
directly measured quantity (the measurements are made on a $10^4$
lattice).}
\end{center}
\end{table}

We measured the two helicity moduli (at flux $0$ and $\pi$) at
$\beta=\beta_c^+$, in the Coulomb phase, on a $32^4$ lattice,
obtaining $h_{0,+}=0.4325(15)$ and $h_{\pi,+}=-1.413(16)$ respectively.
Then, we extracted the renormalized coupling from
Eqs.~(\ref{eq:hm_and_renorm_coupling}) and (\ref{eq:hm_and_renorm_coupling_at_flux_pi}).
This gives

\begin{eqnarray}
e^2_{R} &=& \frac{1}{\beta_{R}}=2.30(1)\quad {\rm for ~} \phi=0\\
e^2_{R} &=& \frac{1}{\beta_{R}}=2.31(1)\quad {\rm for ~} \phi=\pi
\end{eqnarray}

\noindent
which are in agreement (on smaller lattices the renormalized coupling at flux zero is systematically smaller than the one at flux $\pi$; one could use this difference as a criterion to evaluate how close to the thermodynamic limit we are.). We take as our
final value the average, with a conservative error sufficient to cover both
determinations: 

\begin{equation}
e^2_R = 2.305(15)
\end{equation}

This value can be compared with those in the literature (up to a
$4\pi$ factor) $e_R^2=2.08(14)$ \cite{Cella:1997hw} and
$e_R^2=2.39(12)$ \cite{Cox:1997wd}. Our value is more accurate.
The main reason is that the other authors compute $e^2_R$ from a
fit of the static potential between two charges, obtained from
Wilson loops measurements; the limitations in the size of the
loops makes the estimate more difficult. In our case, instead (see
Eq.~(\ref{eq:helicity_modulus_spread})), we do not have this
limitation and use the lattice in its whole extension.\\

We can now go back and consider the analysis performed in the
caption of Fig.~\ref{fig:figure4}. The parabolic curve, computed
at flux $\phi=\pm 2\pi$, gives the classical suppression of the
corresponding flux sectors. But a renormalized $\beta$ appears in
front of it: the value $\beta_{\rm{eff.}}=0.85$ was computed
through the requirement that the parabola had to pass through the
local minima. Instead, from a measurement of the helicity modulus
at $\beta=1.2$ one finds $\beta_R=0.852(2)$. This perfect
consistency is yet another check of the validity of
Eqs.(\ref{eq:renorm_near_to_cl_limit},\ref{eq:free_energy_for_fluxes}).

\subsection{Universality of the helicity modulus}
\label{sec:universality_of_the_hm}

In \cite{Cardy:jg} Cardy made a conjecture about the universality of the helicity modulus
(that is, of the renormalized coupling) at the transition point. In the typical
renormalization group language, `universality' means that we are going to find the same
value in all Abelian theories that share the same `long range' properties, when approaching
the boundary with the confined phase from the Coulomb side.

This conjecture is motivated by the formal analogies existing
between the $4d$ compact $U(1)$ theory and the $2d$ XY model. Let
us start by recalling the hamiltonian of the lattice $2d$ XY model

\begin{equation}
{\cal{H}}=-\rho\sum_{\langle ij\rangle} \cos(\theta(r_i) -
\theta(r_j))
\end{equation}

\noindent
where $\rho$ is the coupling parameter (which has
dimension of energy), $\theta(r_i)$ is the spin angle at site
$r_i$, and the sum is over first neighbors. This leads, in the
continuum, to

\begin{equation}
{\cal{H}}=\frac{1}{2}\int d^2x \hspace{0.1cm}\rho \hspace{0.1cm}
(\nabla \theta(\vec x))^2
\end{equation}

\noindent
and in this context $\rho$ is called the \emph{helicity
modulus} (in analogy with what we did in the previous section, one
can argue that there is a difference between the bare $\rho$,
which is just a coupling constant, and the renormalized $\rho_R$,
which
takes into account the fluctuations of the system).\\

The partition functions of both theories admit a decomposition of
the form \cite{Jose:1977gm},\cite{Banks:1977cc}

\begin{equation}
\label{eq:partition_func_decomposition} Z=Z_{P}Z_{I}
\end{equation}

\noindent
where $Z_P$ is the partition function of a gas of
massless excitations (spin waves for the XY model and photons --
hence the subscript $P$ -- for $U(1)$) and of interacting -- hence
the subscript $I$ -- topological excitations (vortices for the XY
model and monopole current loops for $U(1)$) interacting via a
Coulombic potential.

Cardy then develops an analogy between the critical exponent
$\eta$ of the $2d$ XY model, characterizing the falloff of the
spin-spin correlator (in the massless, large-$\beta$ phase)

\begin{equation}
\label{eq:XY_spin_spin_corr} \langle
e^{i\theta(r_1)}e^{-i\theta(r_2)}\rangle \sim \mid r_1-
r_2\mid^{-\eta}
\end{equation}

\noindent
and the renormalized coupling for the Abelian theory:
both quantities arise from the additive renormalization of the
bare coupling by the susceptibility of the topological defects

\begin{eqnarray}
\label{eq:U(1)_XY_analogy1} 2\pi\eta &=&
\frac{1}{\beta}-\pi^2\sum_r(r/a)^2\langle
m(0)m(r)\rangle_I \\
\label{eq:U(1)_XY_analogy2} \frac{1}{\beta^R}\equiv e^2_R
&=&\frac{1}{\beta}-\frac{\pi}{24}\sum_{r,\mu}(r/a)^2\langle
m_\mu(0)m_\mu(r)\rangle_I
\end{eqnarray}

\noindent
where $m$ represents the defect field, $\beta$ is the
bare coupling and the index $I$ in $\langle \cdot \rangle_I$
represents the interacting part $Z_I$ of the partition function in
Eq.~(\ref{eq:partition_func_decomposition}). The crucial step is
to show the universality of $\eta$ in the XY model. Computing
explicitly the left side of Eq.~(\ref{eq:XY_spin_spin_corr})
\cite{Nelson}, one finds that

\begin{equation}
\eta=\frac{k_B T}{2\pi \rho}
\end{equation}

\noindent
where $k_B$ is the Boltzmann constant, which provides a
relation between $\eta$ and the helicity modulus. Taking the limit
$T \to T_c^-$ we find $\eta_c=\frac{1}{4}$: this result is
independent of the specific RG trajectory considered, and
therefore is \emph{universal}, in the sense that it is unchanged
when the short range features of the theory are modified. The
combination of this result with the similarity between
Eqs.~(\ref{eq:U(1)_XY_analogy1}) and (\ref{eq:U(1)_XY_analogy2})
is the basis for Cardy's conjecture that $e^2_R$ is a universal
quantity for the $4d$ Abelian theory.

\begin{figure}[h]
\begin{center}
\includegraphics[height=10.0cm,angle=-90]{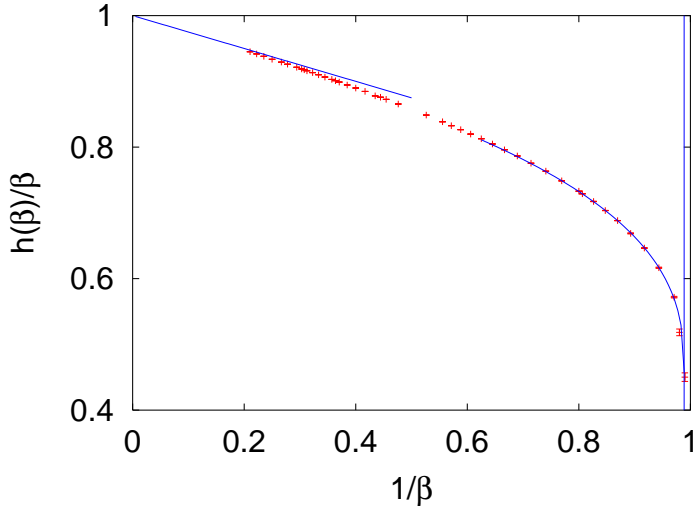}
\end{center}
\caption{\label{fig:helicity_modulus_U1_vs_XY} Helicity modulus in
the $4d$ $U(1)$ theory ($6^4$ lattice), divided by $\beta$ for
direct comparison with the helicity modulus in the $2d$ XY model,
as a function of ``temperature'' $1/\beta$. The
linear and the power law expansions are represented around the
regions $1/\beta\sim 0$ and, respectively, approaching the
transition. The vertical line represents the transition point.}
\end{figure}

The check of this hypothesis can be performed on different levels.
To get a first qualitative idea of the analogy between the two
theories we plot the quantity $\frac{h(\beta)}{\beta}$ of the $4d$
theory (Fig.~\ref{fig:helicity_modulus_U1_vs_XY}), which can be
compared directly with the helicity modulus in the XY model
(our definition of the helicity modulus for $U(1)$ includes an
extra factor $\beta$). In the XY model, the behavior near $T=0$ is
given by $(1-\frac{1}{4}T)$, and approaching $T_c^-$ by a
power-law singularity
($h_c(1+\frac{1}{2}(1-\frac{T}{T_c})^{\nu}+{\cal{O}}(1-\frac{T}{T_c}))$,
$\nu=0.5$) \cite{Nelson}.

Considering the $U(1)$ case, we obtain the same behavior
$\frac{h(\beta)}{\beta} = 1-\frac{1}{4 \beta} + {\cal
O}(\frac{1}{\beta^2})$ from a weak coupling expansion
\footnote{Consider the helicity modulus for the Wilson action
$h(\beta)=\frac{\beta}{V}\langle\sum
\cos\theta_P\rangle-\frac{\beta^2}{V}\langle(\sum
\sin\theta_P)^2\rangle$; one  can show \cite{Horsley:1981gj} that
the first term $\beta\langle
\cos\theta\rangle=\beta-\frac{1}{4}+{\cal O}(\frac{1}{\beta})$,
while the second term turns out to be ${\cal O}(\frac{1}{\beta})$.
A generalization to the extended action is straightforward.}.
Secondly, we measure the exponent $\nu$ from the ansatz

\begin{equation}
\label{eq:critical_behavior_hm}
\frac{h(\beta)}{\beta}=\frac{h(\beta_c^+)}{\beta_c}(1+\alpha(1-\frac{\beta_c}{\beta})^\nu)
\end{equation}

\noindent
close to the transition. The resulting exponent grows systematically with the
lattice size, as shown in Tab.~\ref{tab:nu_vs_L}. From a full jackknife analysis of the
data on our largest lattice ($L=12$), we obtain $\nu=0.34(3)$. This value compares well
with the value $\nu=0.43\pm 0.10$ in \cite{Cella:1997hw}, computed using the same ansatz
for the renormalized coupling data (extracted from Wilson loops) in the vicinity of
$\beta_c$. It is not excluded that the value $0.5$ is attained, as in the XY model, in the
thermodynamic limit.
\begin{table}
\begin{center}
\begin{tabular}{|c|c|}\hline

$L$  & $\nu$ \\

\hline

8 &   0.28(2)   \\ \hline

10 &  0.30(1)  \\ \hline

12 &  0.34(3)  \\ \hline

\end{tabular}
\caption{\label{tab:nu_vs_L}
Fitted exponent $\nu$ from Eq.~(\ref{eq:critical_behavior_hm}),
for increasing lattice size $L$.}
\end{center}
\end{table}
The overall agreement between the helicity modulus in the two
theories is remarkable, and poses the issue of the universality of
the jump of the order parameter at the transition on a more
intuitive basis.\\

\begin{figure}[h]
\begin{center}
\includegraphics[height=11.0cm,angle=-90.]{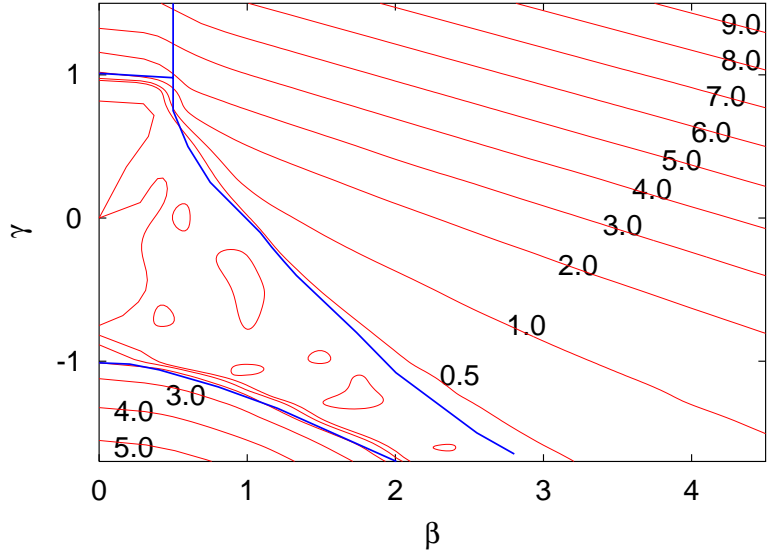}
\end{center}
\caption{\label{fig:iso_helicity_modulus} Lines of constant physics (i.e. constant
renormalized Coulomb coupling $\beta_R = 1/e_R^2$) in the extended $(\beta,\gamma)$
bare coupling plane.
The numbers indicate the value of the helicity modulus ($\approx \beta_R$)
along the various iso-lines.
Solid blue lines indicate the phase boundaries. In the confined phase surrounding the origin,
the renormalized $\beta_R$ is zero.}
\end{figure}

Going to a more quantitative footing, a first possible test is to
compare the values of the renormalized coupling one obtains using
different lattice discretizations of the continuum Abelian theory.
In \cite{Cella:1997hw} a comparison is performed between the
Wilson and the Villain action, which shows agreement of
$e_R^2(\beta_c^+)$ within errors, thus supporting Cardy's
conjecture.

Here we propose a more demanding  test: if universality holds, the
value of the helicity modulus should be the same along the phase
boundary in the extended phase plane. This statement can be
motivated by observing that each phase is actually characterized
by some specific infra-red (IR) behavior, and this is all we need
to apply the universality hypothesis. Therefore, in order to
verify Cardy's conjecture, we studied the \emph{lines of constant
physics} in the Coulomb phase of our model, i.e., the lines of
equal renormalized coupling. They are shown in
Fig.~\ref{fig:iso_helicity_modulus}.

\begin{figure}[h]
\begin{center}
\includegraphics[height=11.0cm,angle=-90.]{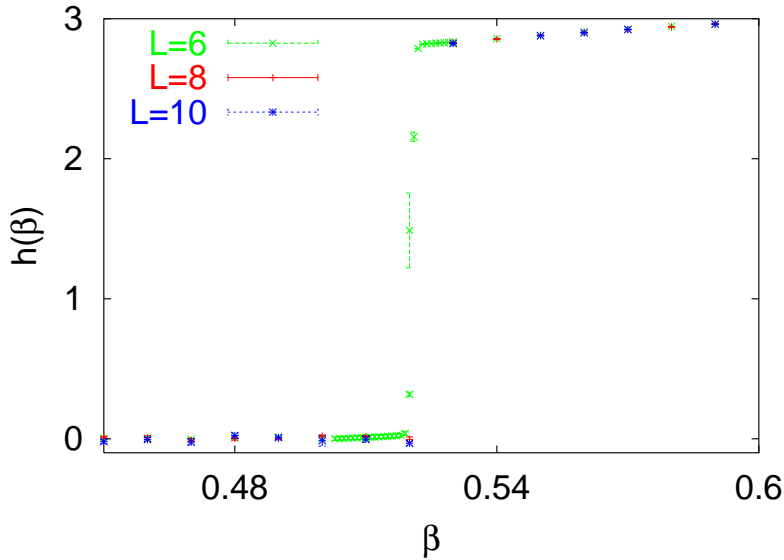}
\end{center}
\caption{\label{fig:hm_gamma_0.9} Helicity modulus versus $\beta$, at $\gamma=0.9$
($6^4,8^4,10^4$ lattice). }
\end{figure}

From this figure we can deduce that Cardy's conjecture is not
satisfied. On one hand, in the lower right corner of the phase
diagram the lines of constant physics tend to be parallel to the
phase boundary, so numerically we should find that the conjecture
is roughly verified (in fact a numerical test performed at
$\beta=2.5$ (Fig.\ref{fig:hm_ext_plane}) gives a value of the
helicity modulus $h=0.37(4)$ in agreement with the Wilson axis
value Eq.~(\ref{eq:fit_results})). On the other hand, if we look at
values of $\gamma>0$, it is clear that the helicity modulus grows,
thus disproving the conjecture. In order to exhibit this effect,
in Fig.~\ref{fig:hm_gamma_0.9} we show the helicity modulus as a
function of $\beta$, for $\gamma=0.9$: it takes value $\sim 2$
near the transition, totally incompatible with the Wilson axis
value of Section \ref{sec:order_of_the_phase}.

As a further comment on Fig.~\ref{fig:iso_helicity_modulus}, it is
interesting to observe that the iso-lines, away from the
boundaries, are roughly straight with slope $-\frac{1}{4}$ (in
Fig.~\ref{fig:iso_helicity_modulus} this is realized in the upper
right corner): this slope  is easily understood considering the
helicity modulus $h(\beta,\gamma)$ at weak coupling, where given a
value $h(\beta,\gamma)=\tilde h$ we find

\begin{equation}
\beta+4\gamma=\tilde h
\end{equation}


\section{Making monopoles with fluxes}
\label{sec:C-periodic_b_c}

Since we have been applying external electromagnetic fluxes to our
system, we consider the natural possibility of enforcing the
presence of a monopole in the lattice through its flux. Such a
possibility is provided by adding an external flux (just like we
did in Section \ref{sec:Charact_of_phases}) but using different
boundary conditions, namely \emph{spatial C-periodic
b.c.}\cite{Davis:2000kv}: when we cross the boundary of the
lattice moving in a spatial direction we enter another copy of the
lattice whose links are all conjugated

\begin{equation}
U_\mu(x+L_i \hat e_i)=U^*_\mu(x),\qquad i=1,2,3,\quad
\mu=1,\ldots,4
\end{equation}

\noindent
The conjugation produces the inversion of the
orientation of the plaquettes, and also the inversion of the
direction of magnetic fluxes. If we  try to vary the flux in such
a lattice, and we insist on requiring translation invariance, we
are faced with a severe restriction: because of the inversion of
the flux direction, on the boundaries we need to impose the
consistency condition

\begin{equation}
\label{eq:flux condition} \phi = - \phi \hspace{0.1cm}
mod\hspace{0.1cm} 2\pi
\end{equation}

\noindent
so only a flux $\phi=0$ or $\pi \hspace{0.1cm} mod
\hspace{0.1cm} 2\pi$ can be added. In the latter case the total
flux going out of the lattice is $2\pi$, and therefore the
presence of one magnetic monopole is enforced in the lattice. What
we called `flux free energy' before can now be rephrased as
\emph{monopole free energy}, and we set

\begin{equation}
F(\phi=\pi,\beta)-F(\phi=0,\beta)=L_4 F_{\rm{monop.}}(\beta)
\end{equation}

\noindent
i.e., $F_{\rm{monop.}}$ is the free energy per
time-slice, or, in more suggestive terms, the free energy per unit
of monopole current.

\begin{figure}[h]
\begin{center}
\includegraphics[height=10cm,angle=-90]{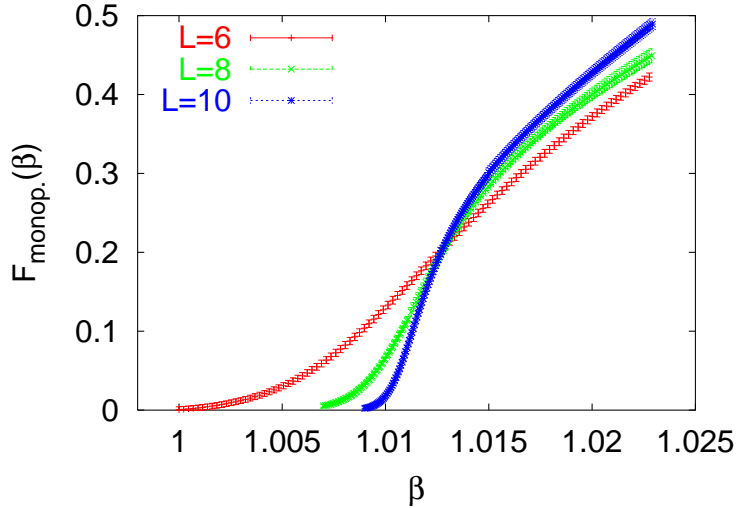}
\end{center}
\caption{\label{fig:rew_monop_free_energy} Monopole free energy as
a function of $\beta$ for different lattice sizes ($L=6,8,10$).}
\end{figure}

In Fig.~\ref{fig:rew_monop_free_energy} we present the numerical
determination of this quantity along the Wilson axis. First of
all, observe that it is an order parameter, as expected: the
monopole free energy is zero in the confined phase and different
from zero in the Coulomb phase. The finite size effects are
clearly visible, as for the helicity modulus. In the Coulomb
phase, as we derived in the previous section, we have finite size
effects proportional to $1/L$ (at leading order). The main
qualitative difference w.r.t. the helicity modulus is that these
lines intersect each other (at $\beta\sim 1.0125$). We will
comment about this during the data analysis.

\begin{figure}[h]
\centerline{
\includegraphics[height=8.2cm,angle=-90.]{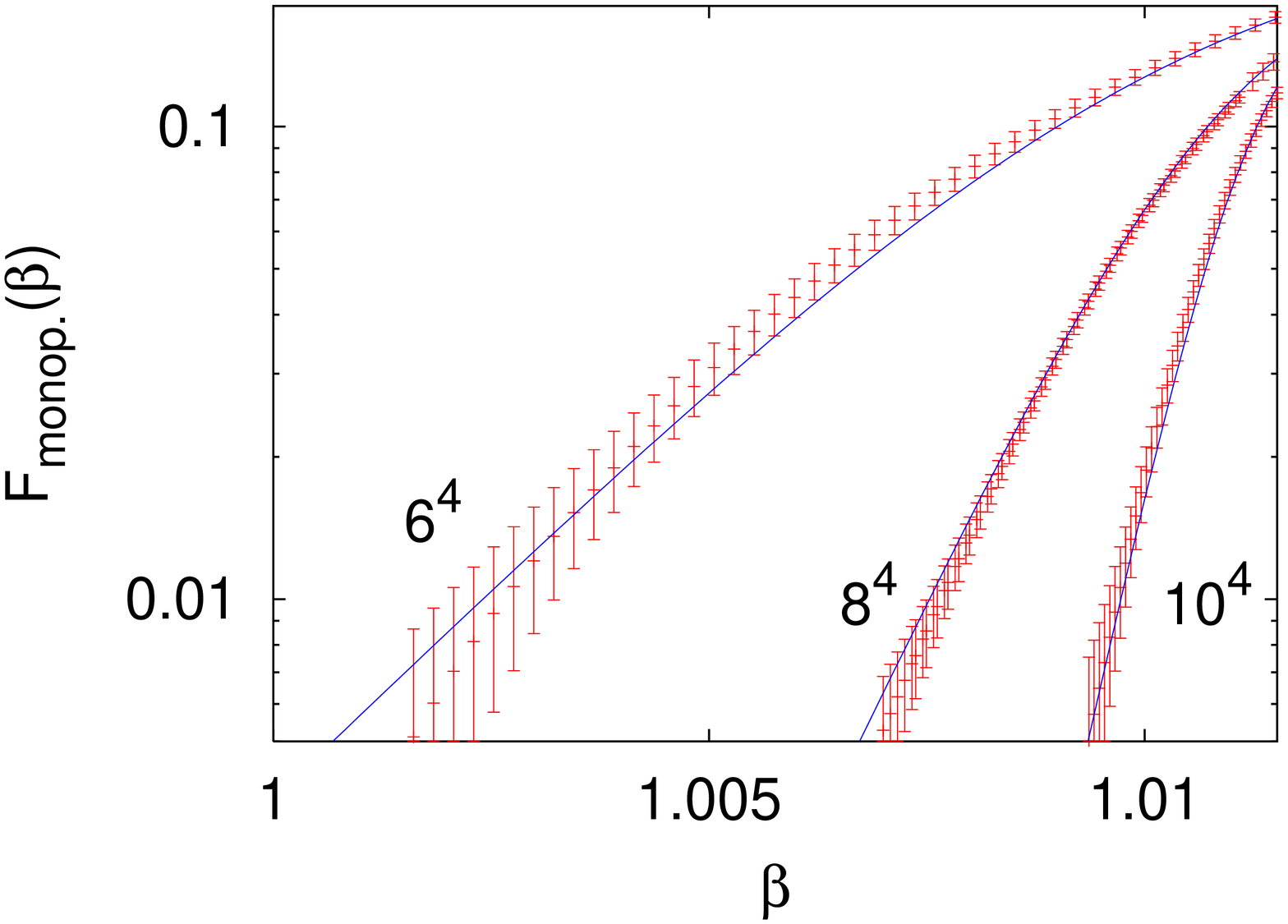}
\includegraphics[height=8.2cm,angle=-90.]{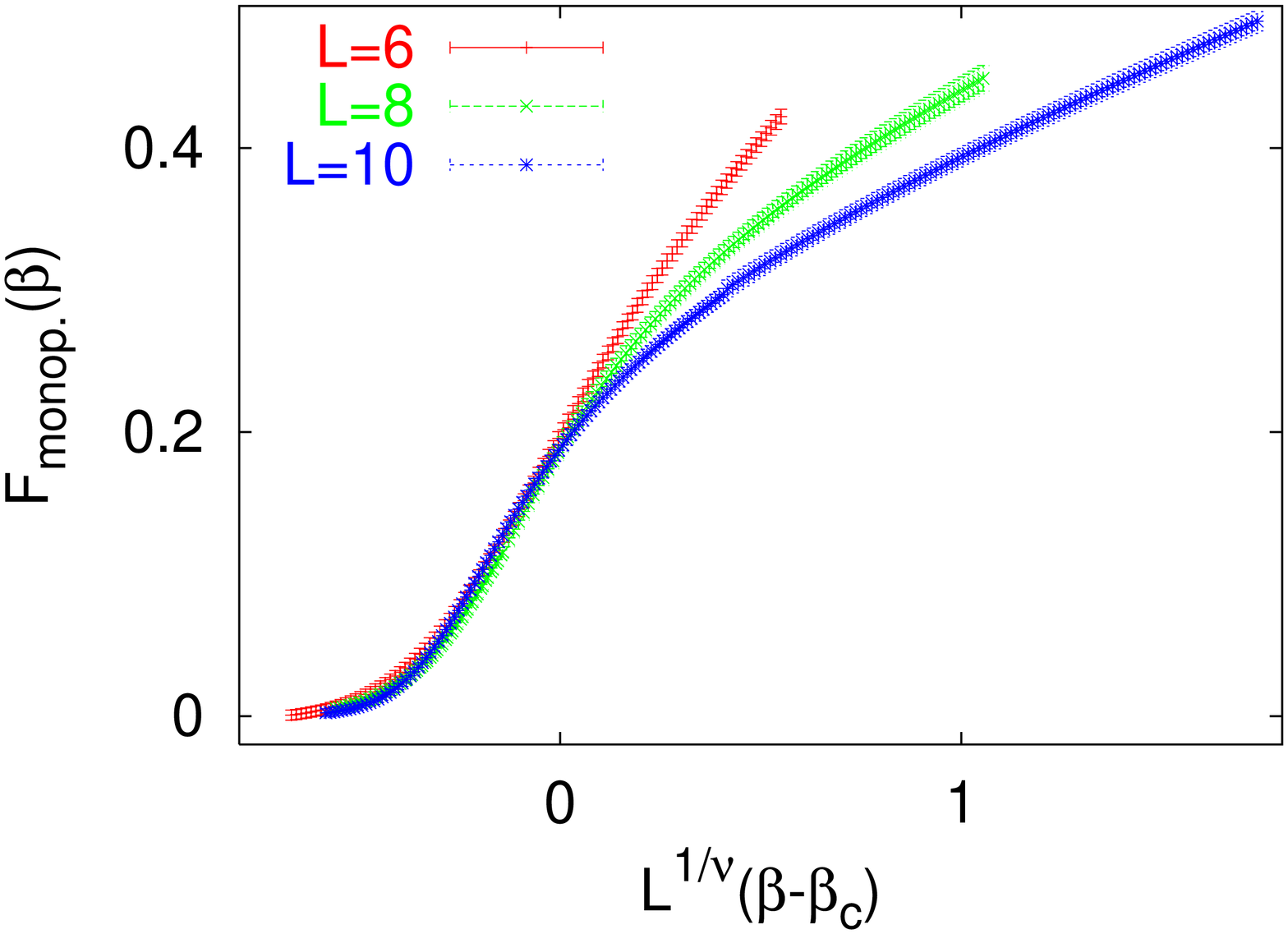}
}
\caption{\label{fig:rew_monop_free_energy_fiest_second_order}
Comparison of the first (left) and second (right) order scenarios for the
monopole free energy data.}
\end{figure}

We now make the same kind of analysis we performed for the
helicity modulus: first we assume that the transition is
first-order, therefore we fit the data of the monopole free energy
in the range $\beta\leq \beta_c$, according to
Eq.~(\ref{eq:hm_first_order_ansatz}). The results are: $m_+ =
0.08(2), \Delta f'/6 = 0.028(5), \log X =0.50(8), \beta_c =
1.0115(3), \alpha_1 \simeq 14.7, \alpha_2 \simeq (5.7)^4 $. In
Fig.~\ref{fig:rew_monop_free_energy_fiest_second_order}, on the
left, we show the result of the fit: the agreement is excellent,
and we can also observe the approximately straight behavior (in
log-scale) of the order parameter away from $\beta_c$, signaling
the first-order nature of the transition.

Going to the second-order analysis, we observe that the data show
good collapse for $\beta < \beta_c$ only (see
Fig.~\ref{fig:rew_monop_free_energy_fiest_second_order} on the
right) when $\beta$ is rescaled according to
Eq.~(\ref{eq:rescale_beta_second_order}) for a `test' value of
$\nu \sim 0.45$. We also observe that the crossing of the curves
of an order parameter at the transition is most unusual for a
second-order phase transition: the usual behavior is that of a
family of non-intersecting curves approaching the thermodynamic
limit.

Also in this case, therefore, we observe an indication in favor of
the first-order nature of the transition.\\

As a numerical algorithm, the chain rule
(Eq.~(\ref{eq:chain_rule})) was used also in this case. It is
interesting to note that all the individual ratios violate the
boundary consistency condition Eq.~(\ref{eq:flux condition}) we
mentioned at the beginning; only the product of all of them
(corresponding to a flux of $\pi$) respects it. To obtain the free
energy over a range of $\beta$ values, we simply apply the
Ferrenberg-Swendsen reweighting algorithm to each ratio of the
chain and then perform
the multiplication.\\

We  conclude this section with some notes about other strategies
used in the literature to enforce the presence of a monopole in
the lattice.

The Pisa group \cite{DiGiacomo:1997sm} builds a magnetically
charged operator $\mu$ which satisfies the following equation

\begin{equation}
\mu(\vec y,t)\mid\hspace{-0.1cm}\vec A(\vec
x,t)\rangle=\hspace{0.1cm} \mid\hspace{-0.1cm}\vec A(\vec
x,t)+\frac{1}{e}\vec b(\vec x -\vec y)\rangle
\end{equation}

\noindent
where $\vec A(x)$ is the vector potential and $\vec b(\vec x-\vec y)$ is the
field of a monopole of charge $\frac{2\pi}{e}$
sitting at $\vec y$; $\mu(\vec y, t)$ can therefore be written as

\begin{equation}
\label{eq:di_giacomo_disorder_parameter} \mu(\vec
y,t)=e^{\frac{i}{e}\int d^3x \vec E(\vec x,t)\vec b(\vec x-\vec
y)}
\end{equation}

The actual \emph{disorder} parameter is built considering a
discretized version of
Eq.~(\ref{eq:di_giacomo_disorder_parameter}). Then  the two-point
function

\begin{equation}
D(t)=\langle \mu(\vec x, t)\bar\mu(\vec x,0)\rangle
\end{equation}

\noindent
describes the propagation of a monopole in time from $0$
to $t$. Finally, considering the usual ansatz

\begin{equation}
D(t)=c e^{-M t}+\langle \mu \rangle^2
\end{equation}

\noindent
one focuses on the measurement of $\langle \mu \rangle$
as a function of $\beta$, which should provide the signature for
monopole condensation.

The boundary conditions used are pure periodic in the spatial
directions and $C$-periodic in time direction. Given a time-slice,
the total flux going out from the boundaries must be zero. This
means that when a monopole is added by hand at position $\vec y$,
an anti-monopole is also automatically created somewhere. In our
case it is possible to enforce the presence of only one monopole
in the lattice because of the $C-$periodic spatial boundary
conditions. A second remark attains the property of translation
invariance: in our case no input about the location of the
monopole in the system is given; this choice is taken dynamically
by the system.

\begin{figure}[h]
\begin{center}
\includegraphics[height=10cm,angle=-90]{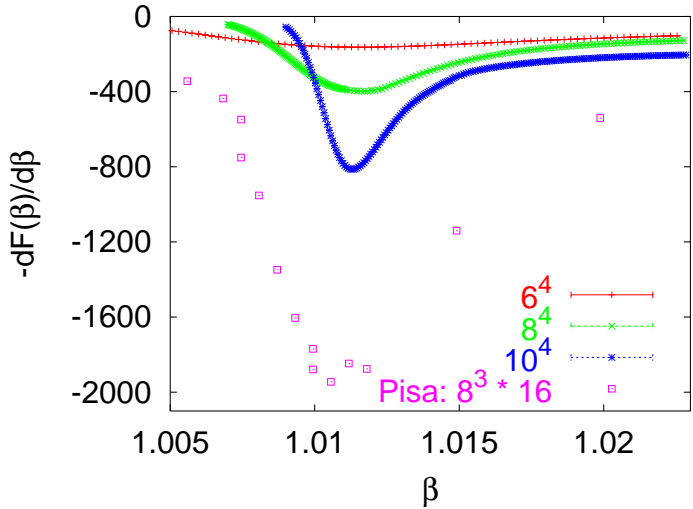}
\end{center}
\caption{\label{fig:compare_digiacomo} Comparison of our data (the
first derivative w.r.t. $\beta$ of the monopole free energy shown
in Fig.~\ref{fig:rew_monop_free_energy}) with Di Giacomo et al.
\cite{DiGiacomo:1997sm}.}
\end{figure}

In order now to compare quantitatively the two constructions, we
show in Fig.~\ref{fig:compare_digiacomo} the derivative w.r.t.
$\beta$ of the monopole free energy, which is the quantity that
directly compares to the Pisa group data. It is interesting to
observe that, for lattices of comparable size, their peak is
roughly twice as high than  ours, which is compatible with the
presence of a pair of monopoles instead of only one. Furthermore,
the width of the peak is also much larger, probably as an effect
of the discretization of the monopole field.

\begin{figure}[h]
\begin{center}
\includegraphics[height=10cm,angle=-90]{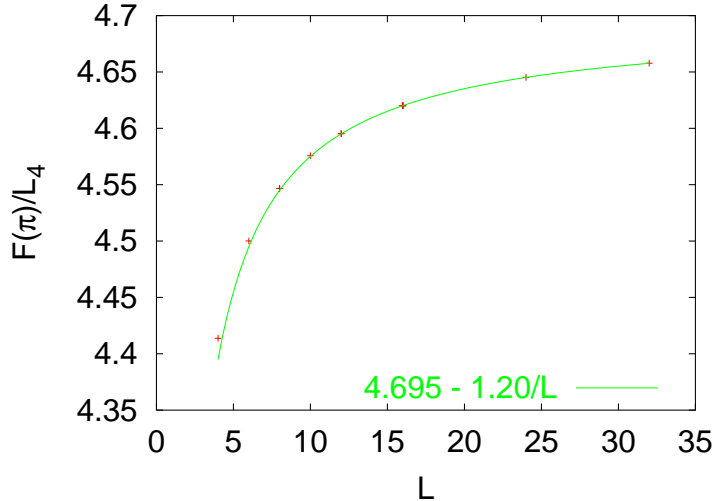}
\end{center}
\caption{\label{fig:class_mon_free_energy} Lattice measurement of
the classical monopole free energy.}
\end{figure}

We can also compare the value of the monopole free energy in the
classical limit. In \cite{DelDebbio:sx} a  computation of the
classical monopole free energy in the weak coupling approximation
is performed. We numerically computed the same quantity by just
cooling the system (limit $\beta \to \infty$, obtained by
minimizing iteratively the action while visiting the links) and
measuring the classical monopole \emph{free energy per time-slice}
$S_{\rm{cl.}}/L_4$: this is the relevant quantity we compare. Our
results are shown in Fig.~\ref{fig:class_mon_free_energy} for the
Wilson action. We obtain

\begin{equation}
\label{eq:classical_monopole_action} S_{\rm{cl}}/L_4 = 4.695
-\frac{1.20}{L}
\end{equation}

\noindent
Ref.\cite{DiGiacomo:1997sm}, Eq.~(58), provides an
expression corresponding to our $(-S_{\rm{cl}})$ in the weak
coupling approximation. Since the lattice there is of size
$L^3\cdot 2L$, we consider

\begin{equation}
\label{eq:cl_mon_frEn_diGiacomo} S_{\rm{cl.}}/L_4=\frac{-(-5.05 L
+ 4.771)}{L_4}=2.525-\frac{2.386}{L}
\end{equation}

\noindent
The rather large difference in the constant terms
between Eq.~(\ref{eq:cl_mon_frEn_diGiacomo}) and
Eq.~(\ref{eq:classical_monopole_action}) should come as no
surprise. We assign it to the discretization of the monopole field
used in the Pisa group computation: since the monopole energy is
UV-divergent, small changes in the monopole field at short
distance will have a large impact. Interestingly, the $1/L$ terms,
which are insensitive to the monopole field discretization, are
very nearly in the ratio 1:2, again in agreement with our
interpretation of the two constructions. \\

Jersak et. al. \cite{Jersak:1999nv} perform an \emph{exact duality
transformation} of the Villain action, imposing spatial
$C-$periodic boundary conditions in the spatial direction and pure
periodic in the time direction. Thus their system is physically
identical to ours. They compute the two-point monopole correlation
functions in the time direction $D(t)$ and use the ansatz

\begin{equation}
D(t)=c e^{-Mt}+\langle\mu\rangle^2
\end{equation}

\noindent
to extract the monopole mass $M$ and monopole condensate
$\langle \mu\rangle$. They measure the monopole condensate as a
function of $\beta$ in the confined region and the monopole mass
(again vs. $\beta$) in the Coulomb phase. Assuming a second order
scenario for the phase transition, a scaling behavior is observed
for both quantities as they approach the (conjectured) critical
point. First of all, it is interesting to observe that in
Fig.~\ref{fig:rew_monop_free_energy_fiest_second_order} (on the
right) we find a `test' critical exponent $\nu=0.45$ which
compares very well with the value they find $\nu=0.49(4)$ for the
critical behavior of the monopole mass in the Coulomb phase
(Eq.~$(1.2)$ in \cite{Jersak:1999nv}). Moreover,  from
Fig.~\ref{fig:rew_monop_free_energy_fiest_second_order} (on the
left) we deduce a value of the monopole free energy at the
transition of about $0.1$, which is smaller than the smallest
value $\sim 0.3$ computed in \cite{Jersak:1999nv} just above the
`critical' point. So Ref.~\cite{Jersak:1999nv} was not in a
position to see the jump of the monopole mass.

\section{Conclusions and perspectives}

In this paper, we have followed 't Hooft's viewpoint that twist
(or flux) free energies can serve as order parameters for confinement.
Twisted boundary conditions were introduced for non-Abelian $SU(N)$
gauge theories. There, the amount of twist is quantized according to
the center group $Z_N$. Here, we have considered the Abelian $U(1)$ theory,
and the twist can be continuously varied. Twisted b.c. are equivalent
to periodic b.c. plus an external electromagnetic flux; we measure
the free energy of such a flux and show that it is an order parameter,
which vanishes in the confined phase just like in the Yang-Mills case
\cite{Kovacs:2000sy,deForcrand:2000fi,DeForcrand:2001dp}. \\

Studying the $U(1)$ theory from the perspective of fluxes allows us
to make practical and conceptual progress. \\

On the practical side, we have introduced the helicity modulus for a
gauge theory, which represents the response of the system to an infinitesimal external flux.
This observable is an order parameter just like the flux free energy.
It is simple to measure, and gives a clear signal for the phase transition
even on a small lattice, because it makes use of the full extent of the lattice.
Moreover, it is simply related to the renormalized coupling.
Using this observable, we show that the transition is first-order,
with parameters ($\beta_c$, latent heat, phase asymmetry) in full agreement
with the recent results of \cite{Arnold:2000hf,Arnold:2002jk}.
Note that a jump in a non-local order parameter is not, by itself, the signature
of a first-order transition. For example, the helicity modulus of the XY model has a
(universal) jump at $T_c$, although the transition is continuous.
The first-order signature comes from the exponential (not power-law)
behavior of the finite-size effects, as seen in Fig.~\ref{fig:hm_rew_fit_quality}, left.

From the helicity modulus we extract the renormalized charge $e_R^2 = 2.305(15)$
at the transition point with high precision. We map lines of constant
renormalized charge in the $(\beta,\gamma)$ coupling plane, and disprove
the conjecture by Cardy \cite{Cardy:jg} of the universality of $e_R$ along the
transition line. \\

On the conceptual side, we have shown that a consistent description of the $U(1)$
phase transition can be given without refering to magnetic monopoles, using
the notion of flux only. In fact, we can build a monopole from flux, and force
a monopole into the system by adjusting the boundary conditions to produce
a $2\pi$ outgoing flux, preserving gauge and translation invariance.
We compare this elegant approach with earlier constructions.

The vanishing of the monopole free energy in the confined phase is thus
accompanied by the vanishing of the flux free energy. Flux sectors,
characteristic of the Coulomb phase, disappear. The condensation of
monopoles is accompanied by the condensation of vortices or Dirac sheets.

From our study, it is clear that monopoles are not the only choice of
topological defect to describe the phase transition. Fluxes and monopoles are
to each other like ``chicken and egg'', and assigning preeminence to one or
the other is a subjective matter in the $U(1)$ case. In the Yang-Mills case,
fluxes are more commonly called center vortices, and monopoles become
Abelian monopoles after Abelian gauge-fixing and projection.
Again they are similarly related to each other. This time, the center
vortices may be given some preference, because their free energy is a
gauge-invariant, UV-regular order parameter \cite{deForcrand:2000fi,DeForcrand:2001dp}. \\

Several extensions of our work come to mind.
The dimension of the system could be reduced to 3. The helicity modulus is
still an order parameter, but at zero temperature it is always zero, 
since the theory is always confining \cite{Polyakov:1976fu}.
At finite temperature however, a deconfinement transition occurs \cite{3dU1}.
This transition is difficult to study numerically, because according to the 
Svetitsky-Yaffe conjecture \cite{SY} it belongs to the Kosterlitz-Thouless universality 
class and is of infinite order. The $U(1)$ helicity modulus, which in this case
directly maps onto the XY helicity modulus, would be a very appropriate order 
parameter.
Another direction would be to measure the helicity modulus in a Yang-Mills
theory after Abelian projection. It may give information about the
deconfinement transition as the temperature is varied \cite{LAT03}.

Finally, the picture in terms of fluxes is the natural environment
to introduce a $U(1)$ \emph{topological charge}. Let
$\Phi_{\mu\nu}$ be the flux through the orientations $\mu,\nu$,
and define $\Phi_{0i}=k_i$, and $\epsilon_{ijk}\Phi_{jk}=m_i$. The
topological charge $Q$ can be expressed as

\begin{equation}
Q=k\cdot m
\end{equation}

\noindent
Even though instantons do not exist in the $U(1)$ theory on $R^4$,
the topological charge above, defined as the integrated second Chern character,
can be non-zero on a torus in the presence of fluxes.\footnote{We thank
O. Jahn and P. van Baal for clarifying this point.} This would
provide a natural laboratory to study the topological charge susceptibility or
$\theta$-vacua \cite{'tHooft:1981ht}, at least in the Coulomb phase.

\section{ACKNOWLEDGEMENTS}
We gratefully acknowledge J\"urg Fr\"ohlich and Oliver Jahn for useful
comments and discussions.


\newpage
\appendix
\section*{APPENDIX: Summary of the simulations performed}
\begin{table}[h]
\begin{center}
\begin{tabular}{|c|c|c|c|c|}\hline

$L$ & $\beta$ &
{\rm meas.} & {\rm update} & {\rm observable} \\

\hline

4  & $(1.2,0)$    & $10^5$  & {\rm multi-canonical} & {\rm flux distribution} \\ 
4  & $(0.8,0)$    & $10^5$  & {\rm multi-canonical} & {\rm flux distribution} \\ 
4  & $(0.8,0)$    & $10^4$  & {\rm snake update}  & {\rm flux free energy}\\ 
4  & $(1.1,0)$    & $10^4$  & {\rm snake update}  & {\rm flux free energy}\\ 
4  & $(1.5,0)$    & $10^4$  & {\rm snake update}  & {\rm flux free energy}\\ 
4  & $(1.0,-1.5)$ & $10^4$  & {\rm snake update}  & {\rm flux free energy}\\ \hline
6  & $(0.99\div 1.03,0)$   &  $10^5$  & {\rm $1$ h.b., $4$ o.r.} & {\rm helicity modulus} \\ 
6  & $(0.99\div 1.03,0)$   &  $10^5$  & {\rm snake update} & {\rm monopole free energy} \\ 
6  & $(1.03\div 1.30,0)$   &  $10^5$  & {\rm $1$ h.b., $4$ o.r.} & {\rm helicity modulus}\\ 
6  & $(2.5,-0.9\div -2.4)$ &  $10^5$  & {\rm $1$ h.b., $4$ o.r.} & {\rm helicity modulus}\\\hline 
8  & $(1.0 \div 1.03,0)$   &  $10^5$  & {\rm $1$ h.b., $4$ o.r.} & {\rm helicity modulus}\\ 
8  & $(1.0 \div 1.03,0)$   &  $10^5$  & {\rm snake update} & {\rm monopole free energy} \\ 
8  & $(2.5,-0.9\div -2.4)$ &  $10^5$  & {\rm $1$ h.b., $4$ o.r.} & {\rm helicity modulus}\\ \hline
10 & $(1.005 \div 1.03,0)$   &  $10^5$  & {\rm $1$ h.b., $4$ o.r.} & {\rm helicity modulus} \\ 
10 & $(1.005 \div 1.03,0)$   &  $10^5$  & {\rm snake update} & {\rm monopole free energy} \\ 
10 & $(1.03 \div 1.30,0)$   &  $10^5$  & {\rm $1$ h.b., $4$ o.r.} & {\rm helicity modulus}\\ 
10 & $(2.5,-0.9\div -2.4)$ &  $10^5$  & {\rm $1$ h.b., $4$ o.r.} & {\rm helicity modulus}\\ \hline
12 & $(1.008 \div 1.03,0)$   &  $10^5$  & {\rm $1$ h.b., $4$ o.r.} & {\rm helicity modulus}  \\ 
12 & $(2.5,-0.9\div -2.4)$ &  $10^5$  & {\rm $1$ h.b., $4$ o.r.} & {\rm helicity modulus}\\ \hline
32 & $(1.0111331,0)$   &  $10^5$  & {\rm $1$ h.b., $4$ o.r.} &  {\rm helicity modulus}\\ \hline

\end{tabular}
\vspace{0.2cm }
\caption{\label{tab:simulation_road_map} 
Summary of the numerical simulations performed 
for this paper. The four columns contain respectively the lattice size $L$, 
the Wilson and extended couplings $(\beta,\gamma)$ (when two values of $\beta$ or $\gamma$ 
are indicated, e.g. $(0.99\div 1.02,0)$, a grid of points is meant in between), 
the number of measurements, the updating algorithm (`h.b.' indicates heat-bath, `o.r.' 
over-relaxation; for the `snake update' \cite{deForcrand:2000fi}, 
see Sec.\ref{sec:Charact_of_phases}), the main observable.\vspace{-0.4cm}}
\end{center}
\end{table}

\end{document}